 \documentclass{JHEP3}
\usepackage{amsmath,amssymb}
\usepackage{graphicx}
\usepackage{cite}
\usepackage{epsfig}
\usepackage{subfigure}
\usepackage{multirow} 

\def\gsim{\ \rlap{\raise 3pt \hbox{$>$}}{\lower 3pt \hbox{$\sim$}}\ }
 \def\lsim{\ \rlap{\raise 3pt \hbox{$<$}}{\lower 3pt \hbox{$\sim$}}\ }

\newcommand{\be}{\begin{equation}}
\newcommand{\ee}{\end{equation}}
\newcommand{\bea}{\begin{eqnarray}}
\newcommand{\eea}{\end{eqnarray}}

\newcommand{\Eqn}[1]{Eq.~(\ref{#1})}

\newcommand{\Eqns}[2]{Eqs.~(\ref{#1})-(\ref{#2})}
 \newcommand{\fh}{f_{H_u}^{eq}}
 \newcommand{\fht}{f_{\widetilde H_u}^{eq}}
\newcommand{\fLeq}{f_{\ell}^{eq}} 
\newcommand{\fLteq}{f_{\widetilde{\ell}}^{eq}}

\newcommand{\muh}{\mu_{\scriptscriptstyle\tilde H}}
\newcommand{\submuh}{\scriptstyle \mu_{\scriptscriptstyle\tilde H}}

\title{
 \vspace{-4pt}
Early Universe effective theories: 
The soft leptogenesis and R-genesis cases
 \vspace{-6pt}
}

\author{Chee Sheng Fong  \\ 
C.N. Yang Institute for Theoretical Physics\\
  State University of New York at Stony Brook\\
  Stony Brook, NY 11794-3840, USA.\\
  E-mail: \email{fong@insti.physics.sunysb.edu}}
\author{M.~C.~Gonzalez-Garcia\\
  Instituci\'o Catalana de Recerca i Estudis Avan\c{c}ats (ICREA), \\
  Departament d'Estructura i Constituents de la Mat\`eria and ICC-UB, \\
  Universitat de Barcelona, 
  Diagonal 647, E-08028 Barcelona, Spain.\\
{\rm and:}  \\ 
C.N. Yang Institute for Theoretical Physics\\
  State University of New York at Stony Brook\\
  Stony Brook, NY 11794-3840, USA.\\
  E-mail: \email{concha@insti.physics.sunysb.edu}}
\author{Enrico Nardi\\
INFN, Laboratori Nazionali di Frascati,\\
Via Enrico Fermi 40, I-00044 Frascati, Italy\\
{\rm and:}  \\ 
Departamento de F\'{\i}sica Te\'orica,  \\
 C-XI, Facultad de Ciencias,  
Universidad Aut\'onoma de Madrid, \\
C.U. Cantoblanco, 28049 Madrid, Spain. \\
{\rm and:}\\  
Instituto de F\'{\i}sica Te\'orica, IFT-UAM/CSIC \\
Nicolas Cabrera 15, C.U. Cantoblanco, 28049 Madrid, Spain\\ 
  E-mail: \email{enrico.nardi@lnf.infn.it}}

\abstract{We discuss the effective theory appropriate for studying
  soft leptogenesis at temperatures $T\gsim 10^{7}\,$GeV.  In this
  regime, the main source of the $B-L$ asymmetry is the CP asymmetry
  of a new anomalous $R$-charge that couples to generalized anomalous
  electroweak processes.  Baryogenesis thus occurs mainly through
  $R$-genesis, and with an efficiency that can be up to two orders of
  magnitude larger than in usual estimates. Contrary to common belief,
  a sizeable baryon asymmetry is generated also when thermal
  corrections to the CP asymmetries in sneutrino decays are neglected
  which, in soft leptogenesis, implies vanishing lepton-flavour CP
  asymmetries.  We present general Boltzmann equations for soft
  leptogenesis that are valid in all temperature regimes.  }


 \preprint{
IFT-UAM/CSIC-10-90
\\
YITP-SB-10-42
}

\begin{document} 

\section{Introduction}
\label{sec:intro}

In the hot and fast expanding Universe, during the first instants
after the Big Bang, at any given temperature $T$ all particle physics
processes having a characteristic time scale $\tau$ larger than the
Universe age $t_U(T)$ do not occur, and must be neglected. This is
important, because generically speaking several particle interactions
that are allowed by the fundamental gauge symmetries violate some
other global conservation laws. However, until the Universe is old
enough that these reactions can occur with rates comparable, or
larger, than the Universe expansion rate, the will-be violated
quantities remain effectively conserved. In the language of field
theory Lagrangians, this means that at each cosmological temperature
$T$, the relevant particle physics processes are determined by an
effective Lagrangian in which all the parameters responsible for
`slow' reactions, that is reactions with characteristic timescales
$\tau \gg t_U(T)$, must be set to zero. By doing this, it is then easy
to identify the new global symmetries of the effective Lagrangian, and
if no anomalies are involved, these symmetries correspond to conserved
quantities.

In the context of the early Universe the meaning of `effective
theory', the one that we will use in this paper, differs somewhat from
what is generally meant in particle physics by `effective field
theory'. The latter case refers to the low energy theory obtained
e.g. from a fundamental Lagrangian when all the states heavier than
some high energy cutoff are integrated out, and corresponds to a
theory with a reduced number of degrees of freedom. In contrast, the
effective theories required to study particle physics processes in the
early Universe correspond to theories with a reduced number of
fundamental parameters.  Let us explain this in some detail: at each
specific temperature $T$, particle reactions must be treated in a
different way depending if their characteristic time scale $\tau$
(given by inverse of their their thermally averaged rates) is
\begin{itemize} \itemsep=-2pt
\item[(i)] much shorter than the age of the Universe:\quad $\tau \ll t_U(T)$; 
\item[(ii)] much larger than the age of the Universe:\quad $\tau \gg t_U(T)$; 
\item[(iii)] comparable with the  Universe age:\quad $\tau \sim t_U(T)$.
\end{itemize}
The first type of reactions (i) occur very frequently during one
expansion time $1/H(T)$ ($H(T)$ being the Hubble parameter at $T$) and
their effects can be simply `resummed' by imposing on the
thermodynamic system the chemical equilibrium condition appropriate
for each specific reaction, that is $\sum_I \mu_I = \sum_F \mu_F$,
where $\mu_I$ denote the chemical potential of an initial state
particle, and $\mu_F$ that of a final state particle. The numerical
values of the parameters that are responsible for these reactions only
determine the precise temperature $T$ when chemical equilibrium is
attained and the resummation of all effects into chemical equilibrium
conditions holds but, apart from this, have no other relevance,and do
not appear explicitly in the effective formulation of the problem.
Reactions of the second type (ii) cannot have any effect on the
system, since they basically do not occur.  Then all physical
processes are blind to the corresponding parameters, that can be set
to zero in the effective Lagrangian.  In most cases (but not in {\it
  all} cases) this results in exact global symmetries that correspond
to conservation laws for the corresponding charges, that must be
respected by the equations describing the dynamics of the system.
Reactions of the third type (iii) in general violate some symmetries,
and thus spoil the corresponding conservation conditions, but are not
fast enough to enforce chemical equilibrium conditions.  Only
reactions of this type appear explicitly in the formulation of the
problem (they generally enter into a set of Boltzmann equations for
the evolution of the system) and only the corresponding parameters
represent fundamental quantities in the specific effective theory.

Several examples of the importance of using the appropriate early
Universe effective theory can be found in leptogenesis studies.
Leptogenesis~\cite{fy,leptoreview}   was first formulated
in the so called `one flavour
approximation'~\cite{Luty:1992un,Covi:1996wh,Buchmuller:1996pa} in
which a single $SU(2)$ lepton doublet of an unspecified flavour is
assumed to couple to the lightest singlet seesaw neutrino, and it is
thus responsible for the generation of the lepton asymmetry.  Until
the works in refs.~\cite{flavour1,flavour2}, most leptogenesis studies
were carried out within this framework, although a few earlier works
had already explored in some detail the effects of lepton flavours in
leptogenesis~\cite{barbieri,morozumi}, or had used them in specific
leptogenesis realizations~\cite{oscar}.

Nowadays, it is well understood that the `one flavour approximation'
gives a rather rough and often unreliable description of leptogenesis
dynamics in the regime when flavour effects are important. This is
because such an `approximation' has no control over the effects that
are neglected, and thus the related uncertainty cannot be estimated.
On the other hand, it is seldom recognized that if leptogenesis occurs
above $T\sim 10^{12}\,$GeV, when all the charged leptons Yukawa
interactions have characteristic time scales much larger than $t_U$,
the `one flavour approximation' is not at all an
approximation. Rather, it is the correct high temperature effective
theory that must be used to compute the baryon asymmetry.  The
corresponding effective Lagrangian is obtained by setting to zero, in
first place, all the charged lepton Yukawa couplings, so that the only
remaining flavour structure is determined by the Yukawa couplings of
the heavy Majorana neutrinos, that must remain non-vanishing in order
that decays into light leptons can occur.

The transition to the regime where the unflavoured effective theory
must be used, corresponds to a different physics framework that is
characterized by an overall reduced amount of CP violation, that is
encoded in a single CP violating parameter $\epsilon$, rather than in
the three (or two, for intermediate temperature
regimes~\cite{flavour1,flavour2,barbieri}) flavoured CP asymmetries
$\epsilon_e,\,\epsilon_\mu,\,\epsilon_\tau$.  Similarly, the lepton
density asymmetry is produced into a single species of leptons rather
than in the three (or two) of the flavoured regimes.  Thus, in the
regime in which the appropriate theory is flavour blind, a lesser
amount of baryon asymmetry can be produced.  Of course, in the same
regime also the rates of other processes, like for example those
induced by the Yukawa couplings of the light quarks~\cite{spectator2}
or the electroweak sphaleron rates~\cite{spectator2,spectator1}, must
be set to zero, but being these typical `spectator' processes, the
effect of switching them off is numerically much less
relevant.\footnote{Spectator processes are fast processes that do not
  violate $B-L$ but that can still have an impact on the baryon
  asymmetry yield of leptogenesis~\cite{spectator1,spectator2}.}

In summary, early works on Standard Model (SM) leptogenesis were
carried out from the start within the unflavoured effective
theory. Quite likely this happened because the corresponding
Lagrangian is much more simple than the full SM Lagrangian given that
the number of relevant parameters is reduced to a few.  The main
virtue of subsequent studies on lepton flavour
effects~\cite{flavour1,flavour2,barbieri,morozumi} was that of
recognizing that below $T\sim 10^{12}\,$GeV the unflavoured theory
breaks down, and the new theory, that at each step the temperature is
decreased brings in new fundamental parameters, can give genuinely
different answers for the amount of baryon asymmetry that is
generated.

In supersymmetric leptogenesis the opposite happened, because the
effective theory that was generally used is in fact only appropriate
for temperatures much lower than the typical temperatures $T\gg
10^8\,$GeV in which leptogenesis can be successful, and only quite
recently it was clarified that in the relevant temperature range a
completely different effective theory holds instead~\cite{ournse}.
More specifically, it was always assumed (often implicitly) that
lepton-slepton reactions like e.g. $\ell\ell \leftrightarrow
\tilde\ell\tilde\ell$ that are induced by soft supersymmetry-breaking
gaugino masses, are in thermal equilibrium (see
refs.~\cite{leptoreview,Plumacher:1997ru,thermal} for examples of well
known papers adopting this assumption). This implies equilibration
between the leptons and sleptons density asymmetries ({\sl
  superequilibration}) while in general, in supersymmetric
leptogenesis, superequilibration (SE) does not occur. In fact,
requiring that the rates induced by supersymmetry-breaking scale
($\Lambda_{susy}$) parameters, like soft breaking masses $\tilde m$ or
the Higgsino mixing parameter $\muh$, are slower than the Universe
expansion rate when $T\sim M$ (being $M$ the heavy neutrino mass, and
$m_P$ below the Planck mass) one obtains
\begin{equation}
  \label{eq:NSE}
  \frac{\Lambda_{susy}^2}{M} \lsim 25\; \frac{M^2}{m_P}
\qquad \Rightarrow \qquad  M \gsim 5\times 10^7
\left(\frac{\Lambda_{susy}}{500\,{\rm GeV}}\right)^{2/3}\,{\rm GeV}. 
\end{equation}
The effective theory appropriate for studying supersymmetric
leptogenesis, in which the heavy Majorana masses certainly satisfy the
bound~\Eqn{eq:NSE}, is thus obtained by setting $\tilde m,\,\muh \to
0$.  The consequences of this were analyzed in~\cite{ournse} and are
far reaching. At $T\gsim  10^{7}\,$GeV, besides the occurrence of
non-superequilibration (NSE) effects, additional anomalous global
symmetries that involve both $SU(2)$ and $SU(3)$ fermion
representations emerge~\cite{Ibanez:1992aj}.  As a consequence, the
electroweak (EW) and QCD sphaleron equilibrium conditions are modified
with respect to the usual ones and, among other things, this also
yields a different pattern of sphaleron induced lepton-flavour
mixing~\cite{barbieri,flavour1,flavour2}. In addition, a new
anomaly-free $R$-symmetry can be defined and the corresponding charge,
being exactly conserved, provides a constraint on the particles
density asymmetries that is not present in the SM.  However, in
\cite{ournse} it was also concluded that, in spite of all these
modifications, the resulting baryon asymmetry would not differ much
from what was obtained in the usual scenario. Basically, this happens
because by dropping the SE assumption and accounting for all the new
effects, only modifies spectator processes, while the overall amount
of CP asymmetry that drives leptogenesis remains the same.

\smallskip    

The most interesting scenario in which the appropriate effective
theory not only yields far reaching qualitative differences but also
very large quantitative effects, is in soft leptogenesis 
(that is leptogenesis where the origin of CP violation is in the soft
supersymmetry-breaking terms~\cite{Boubekeur:2002jn,soft1,soft2}) if
it occurs above the SE threshold~\Eqn{eq:NSE}.  This is because of two
main reasons:

(I)\ \ The first one is that in soft leptogenesis there is a strong
cancellation between the CP asymmetries for sneutrino decays into
scalars and into fermions $\epsilon\equiv \epsilon_s + \epsilon_f
\simeq 0 $. This cancellation is almost exact in the $T=0$ limit, and
$\epsilon$ gets lifted to an appreciable level only when thermal
corrections are included~\cite{soft1,soft2}.  In the NSE regime
however, the independent evolution of the scalar and leptonic density
asymmetries implies that the corresponding efficiencies $\eta_{s,f}$
are different.  When these different `weights' are taken into account, 
the cancellation between the scalar and fermion contributions to the
baryon asymmetry gets spoiled and a non-vanishing result is obtained
even in the $\epsilon\equiv \epsilon_s + \epsilon_f \to 0 $ limit.
Note that this effect can dominate over the ones due to thermal
(or higher order) corrections to the CP asymmetries.  This situation
is reminiscent of the so called Purely Flavoured Leptogenesis (PFL)
scenarios~\cite{flavour2,PFL1,PFL2,Antusch:2009gn} where the vanishing
of the total CP asymmetry resulting from the sum over lepton flavours
$\epsilon=\sum_\alpha\epsilon_\alpha=0$ ($\alpha=e,\mu,\tau$) does not
imply a vanishing baryon asymmetry ${\Delta B}\propto
\sum_\alpha\eta_\alpha\epsilon_\alpha\neq 0$. This, provided that lepton
flavour equilibrating (LFE) reactions $\ell_\alpha \leftrightarrow
\ell_\beta$, that in the PFL case play the same role than SE for soft
leptogenesis, remain out of equilibrium~\cite{lfe}.

(II)\ \ The second reason is even more interesting. In the high
temperature effective theory two new global symmetries (a $R$-symmetry
and a $PQ$-like symmetry) arise. While these symmetries are anomalous,
two new anomaly free combinations of charges involving $R$ and $PQ$
can be defined.  These new charges, that we denote as ${R_B}$ and
${R_\chi}$, are only (slowly) violated by sneutrino dynamics, that is
by reactions of the third type (iii) in the classification given
above, and thus their evolution must be followed by means of two new
Boltzmann Equations (BE).  Because charge density asymmetries get
mixed by EW sphalerons, these equations are coupled to the BE that
control the evolution of the ${B-L}$ asymmetry, and thus the dynamical
evolution of ${R_B}$ and ${R_\chi}$ affects its final value.  What is
important, is that the CP violating sources for these two charges,
that are respectively $\epsilon_s$ and $\epsilon_s-\epsilon_f$, are
not suppressed by any kind of cancellation, and the corresponding
density asymmetries remain large during leptogenesis. They act as
source terms for ${B-L}$ that is thus driven to comparably large
values.  As regards the final values of ${R_B}$ and ${R_\chi}$ at the
end of leptogenesis, they are instead irrelevant for the computation
of the baryon asymmetry since, well before the temperature when the EW
sphalerons are switched off, soft supersymmetry-breaking effects
attain in-equilibrium rates, implying that $R$ and $PQ$ cease to be
good symmetries also at the perturbative level. Thus, they decouple
from the sphaleron processes that then reduce to the usual SM $B-L$
conserving form, that involves only quarks and leptons.  The baryon
asymmetry is then given only by $B-L$ conversion, according to the
usual equation $B=\frac{8}{23}\, (B-L)$.

The outline of the paper is as follows: In
Section~\ref{sec:motivations} we recall the main motivations for soft
leptogenesis and review recent results and the relevant literature.
The soft leptogenesis scenario is summarized in
Section~\ref{sec:softL}: we first present the relevant Lagrangian and
next, in Section~\ref{sec:softCP}, we compute the various CP
asymmetries including some subleading terms for the CP asymmetries
``in mixing'' -- generated by sneutrino self-energy diagrams -- that
avoid a complete cancellation between the fermions and bosons
contributions even in the $T\to 0$ limit.  Conversely, for the CP
asymmetries ``in decays'' -- that are generated by vertex corrections
-- we present in Section~\ref{sec:vertexCP} a simple proof ensuring
that at one loop the cancellation at $T=0$ is exact.  The effective
theory appropriate for studying the generation of the baryon asymmetry
when the heavy sneutrino masses satisfy the bound in~\Eqn{eq:NSE} is
described in Section~\ref{sec:NSE}.  We derive the equilibrium
conditions and the relevant conservation laws that constrain the
particle density asymmetries, we identify the new quasi-conserved
charges, and we also compute the matrices that control the sphaleron
induced lepton flavour mixing for two different sets of values of the
electron and down-quark Yukawa couplings.  In Section~\ref{sec:BE} we
present the set of the {\em five} basic BE, that is valid for
numerical studies of soft leptogenesis at all temperatures, and in
Section~\ref{sec:simple} we discuss a simple case in which the role
played by the $R_B$ and $R_\chi$ charge asymmetries is particularly
transparent.  In Section~\ref{sec:results} we compute numerically the
amount of baryon asymmetry that can be generated in soft leptogenesis
within the NSE regime, and compare it to previous results based on the
assumption of SE. Finally in Section~\ref{sec:conclusions} we present
a simple explanation of the large numerical enhancements, we recap the
main results and draw the conclusions. Two Appendix complete the
paper: in Appendix~\ref{Appendix-A} we collect the relevant thermal
factors, in Appendix~\ref{Appendix-B} we present a more complete set
of BE which also include scatterings.

\section{Soft leptogenesis: review and motivations}
\label{sec:motivations}
With the discovery of neutrino oscillations,
leptogenesis~\cite{fy,leptoreview} became a particularly well
motivated mechanism to explain the cosmic baryon asymmetry. This
happened because the scale of the oscillation mass square differences
is perfectly compatible with sufficient deviations from thermal
equilibrium in the decays of the heavy seesaw
neutrinos~\cite{Fischler:1990gn}.  For a hierarchical spectrum of the
heavy Majorana states, successful leptogenesis requires generically a
quite large leptogenesis scale~\cite{di}, corresponding to seesaw
neutrino masses of order $M>2.4 (0.4)\times 10^9$~GeV for vanishing
(thermal) initial neutrino
densities~\cite{di,Mbound,thermal}.\footnote{These limits are
  basically not affected by flavour effects~\cite{flavour3,db2}.  With
  resonantly enhanced CP asymmetries~\cite{PU} or in various extended
  scenarios~\cite{lowtemp} lower leptogenesis scales are instead 
  possible.}

The presence in the theory of such a
large mass scale poses a serious fine tuning problem for keeping the
Higgs mass parameter at the electroweak scale~\cite{Casas:2004gh}.
Low-energy supersymmetry can be invoked to naturally stabilize the
hierarchy between this new scale and the electroweak one. This,
however, introduces a certain conflict between the gravitino bound on
the reheat temperature and the thermal production of the heavy
singlets neutrinos \cite{gravi}.

Once supersymmetry is introduced, there are, however, new sources of
lepton number and CP violation that are related to the soft
supersymmetry-breaking terms involving the sneutrinos. This allows for
a different leptogenesis scenario that is specific to supersymmetry
known as `soft leptogenesis'~\cite{Boubekeur:2002jn,soft1,soft2}.
Given that the new effects are generically suppressed by powers of the
ratio between the soft supersymmetry-breaking scale and the sneutrino
masses $\Lambda_{susy}/M$, the characteristic temperature window in
which the new contributions can give relevant effects is roughly
$10^{4}-10^{8}$ GeV.  Thus, apart from a small temperature interval
lying approximately within $10^{8}\,{\rm GeV}\lsim T\lsim 10^{9}\,{\rm
  GeV}$, supersymmetric leptogenesis can proceed at any temperature
above the EW scale, and the low temperature soft leptogenesis
realization allows to evade the gravitino problem.  Note however, that
the presence of a forbidden temperature window implies that the two
different leptogenesis realizations never overlap. Thus, when studying
soft leptogenesis, the standard contributions to the CP asymmetries
give irrelevant effects and can be safely neglected, and the opposite
is true when the standard high temperature scenario is assumed.

In the original papers on soft leptogenesis~\cite{soft1,soft2} only
one type of contributions to the CP asymmetries in sneutrino decays
was identified: the so called CP violation in mixing.  CP violation in
mixing is induced by the bilinear sneutrino $B$ term that removes the
mass degeneracy between the two real sneutrino states.  As for the
case of resonant leptogenesis~\cite{PU}, the sneutrino self-energy
contributions to the CP asymmetries can then be resonantly enhanced,
and give rise to sufficiently large CP violation in sneutrino decays.
However, to satisfy the resonant condition, unconventionally small
values of $B$ are required~\cite{soft1,soft2,ourflasoft,ourqbe}.
Extended scenarios were thus proposed in order to alleviate this
problem~\cite{ourinvsoft,softothers}.
However, it was also realized that within the context of the minimal
scenario, besides CP violation in sneutrino mixing additional sources
of CP violation can arise from vertex corrections to the decay
amplitudes, and from the interference between mixing and
decay~\cite{soft3,ourgaugino}.  These new sources of CP violation (the
so called ``new ways to soft leptogenesis''~\cite{soft3}) are induced
by gaugino soft supersymmetry-breaking masses that appear in vertex
corrections to the decays.  With respect to the mixing contributions,
these corrections are suppressed by more powers of $\Lambda_{susy}/M$,
and thus they can be sizable only at relatively low temperatures
$T\lsim 10^5\,$GeV~\cite{soft3}. Although in this regime they can
allow for more conventional values of $B$, such a low seesaw scale
implies that the suppression of the light neutrino masses is mainly
due to very small values of the Yukawa couplings $|h_\nu|^2\sim
10^{-10}$ rather than to the seesaw  scale.

Concerning the role of flavour~\cite{flavour1,flavour2,barbieri%
  ,morozumi,flavour3,db2,flavourothers,riottoqbefla,riottosc,lfe} and
spectator effects~\cite{spectator1,spectator2}, they had all been
neglected in the original soft leptogenesis
papers~\cite{soft1,soft2,soft3} that were based on the single-flavour
scenario.  However, soft leptogenesis always occurs at temperatures
where the appropriate effective theory must include the effects of the
three lepton flavours. This was done in Ref.~\cite{ourflasoft} that
also included spectator effects, but that assumed a constrained
scenario with universal trilinear couplings.  Within this context, it
was found that the leptogenesis efficiency could be enhanced by ${\cal
  O }(30)$ with respect to the single flavour analysis.  The more
general scenario of non-universal trilinear couplings was considered
in~\cite{oursoftnu}, that also included the possibility of damping
flavour effects through large LFE spectator processes~\cite{lfe}. It
was found that when the assumption of universality is dropped, flavour
effects can play an even more important role, with the possibility of
enhancing the leptogenesis efficiency by more than three orders of
magnitude with respect to the one flavour treatment.

In spite of all these advancements and refinements in soft
leptogenesis studies, a crucial point has been always overlooked.  As
was first pointed out in~\cite{ournse}, when the sneutrino masses
satisfy the bound~\Eqn{eq:NSE} the appropriate effective theory for
studying early Universe processes in a supersymmetric scenario is
different from the one that has always been used.  As we will show, in
soft leptogenesis the correct effective theory implies particularly
dramatic effects, namely the final baryon asymmetry that is produced
can be up to two orders of magnitude larger than what is obtained with
the previous treatments.


\section{Soft Leptogenesis Lagrangian and CP asymmetries} 
\label{sec:softL}
The superpotential for the supersymmetric seesaw model is: 
\begin{equation}
W=\frac{1}{2}M_{ij}N^c_{i}N^c_{j}+Y_{i\alpha}
N^c_{i}\ell_{\alpha} H_u ,
\label{eq:superpotential}
\end{equation}
where $i,j=1,2,\dots$ label the chiral superfields of the heavy
$SU(2)$ singlet Majorana neutrinos defined according to usual
conventions in terms of their left-handed Weyl spinor components
($N^c$ has scalar component $\tilde N^*$ and fermion component
$N^c_L$), $\alpha=e,\mu,\tau$ labels the flavour of the $SU(2)$ lepton
doublets $\ell_{\alpha}=\left(\nu_{\alpha},e_{\alpha}^{-}\right)^T$,
$H_u=(H_u^+,H_u^0)^T$ denotes the up-type Higgs doublet superfield,
and contraction of the $SU(2)$ indexes between doublets $\ell_{\alpha}
H_u = \epsilon_{\rho\sigma} \ell_{\alpha}^\rho H_u^\sigma$ with
$\epsilon_{12}=+1$ is left understood.

The relevant soft supersymmetry-breaking terms involving the scalar
components of the $N^c$ superfields  
and the $SU(2)$ gauginos
$\widetilde{\lambda}_{2}^{a}$ are given by
\begin{eqnarray} 
-\mathcal{L}_{soft} & = 
& \widetilde{M}_{ij}^2 \widetilde{N}^*_{i} \widetilde{N}_{j} + 
\left(A Y_{i\alpha}\widetilde{N}^*_{i}
\tilde{\ell}_{\alpha} H_u +\frac{1}{2}B 
M_{ij}\widetilde{N}^*_{i}\widetilde{N}^*_{j}
+\frac{1}{2}m_{2}\overline{\tilde{\lambda}}_{2}^{a}
P_{L}\tilde{\lambda}_{2}^{a} +\mbox{h.c.}\right)\;. 
\label{eq:soft_terms}
\end{eqnarray}
$U(1)$ gauginos can be straightforwardly included in similar form.
In~\Eqn{eq:soft_terms} we have assumed for simplicity universal
trilinear and bilinear couplings $A_{i\alpha} =A Y_{i\alpha}$ and
$B_{ij} =B M_{ij}$.
The sneutrino and anti-sneutrino states mix, resulting in the mass
eigenstates:
\begin{eqnarray}
\widetilde{N}_{+i} & = &
\frac{1}{\sqrt{2}}(e^{i\Phi/2}\widetilde{N}^*_{i}+e^{-i\Phi/2}
\widetilde{N}_{i}),\nonumber
\\ \widetilde{N}_{-i} & = &
\frac{-i}{\sqrt{2}}(e^{i\Phi/2}
\widetilde{N}^*_{i}-e^{-i\Phi/2}\widetilde{N}_{i}),
\label{eq:mass_eigenstates}
\end{eqnarray}
where $\Phi\equiv\arg(BM)$, and 
$\widetilde{N}_{\pm i}$  have mass eigenvalues: 
\begin{eqnarray}
M_{\pm i}^{2} & = & M_{ii}^{2}+\widetilde{M}_{ii}^2 \pm|B M_{ii}|.
\label{eq:mass_eigenvalues}
\end{eqnarray}

The interaction Lagrangian involving the mass
eigenstate sneutrinos $\widetilde{N}_{\pm i}$, the Majorana singlet neutrinos
$N_{i}$, the $SU(2)$ gauginos $\tilde{\lambda}_{2}$ and the (s)leptons
and Higgs(inos) doublets reads:
\begin{eqnarray}
-\mathcal{L}_{int}&=& \frac{Y_{i\alpha}}{\sqrt{2}}
\left[
\widetilde{N}_{+i}\left(\overline{\widetilde{H_u}^c} 
P_{L}\ell_{\alpha} +(A + M_{i})
\widetilde{\ell}_{\alpha} H_u^{\beta}\right) 
+ i \widetilde{N}_{-i}\left(
\overline{\widetilde{H}^c_u} P_{L}\ell_{{\alpha}} 
+(A  -M_{i})
\widetilde{\ell}_{\alpha} H_u\right)
\right] \nonumber \\  
&+& 
  Y_{i\alpha}\left[\,
\overline{\widetilde{H}^c_u} P_{L}N_{i}\widetilde{\ell}_{\alpha}
+  \overline{N}_i P_L\ell_\alpha H_u\right]
\nonumber 
\\
 &&  
\hspace{-1.5cm}
+
  g_{2}\left[
\overline{\widetilde{\lambda}}_{2}^{\pm}P_{L}
\ell_{\alpha}\, \sigma_\pm \widetilde{\ell}_{\alpha}^{*}
+
\overline{\widetilde{\lambda}}_{2}^{\pm}P_{R}
\widetilde{H}_{u}^{c} \sigma_\mp H_u
-\frac{1}{\sqrt{2}}\left(\overline{\widetilde{\lambda}}_{2}^{0}P_{L}
\ell_{\alpha}\sigma_3 \tilde{\ell}_{\alpha}^{*}+
\overline{\widetilde{\lambda}}_{2}^{0}P_{R}
\widetilde{H}_{u}^{c} \sigma_3 H_u
 \right)\right]
\!+\!\mbox{h.c.},
\label{eq:mass_basis}
\end{eqnarray}
where $P_{L,R}$ are respectively the left and right chiral projectors,
$\sigma_\pm=(\sigma_1\pm i\sigma_2)/2$ with $\sigma_{i}$ being the
Pauli matrices, and $SU(2)$ contractions like $\ell_{\alpha}\,
\sigma_\pm\, \widetilde{\ell}_{\alpha}^{*}= \ell_{\alpha}^\rho\,
(\sigma_\pm)_{\rho\sigma}\, \widetilde{\ell}_{\alpha}^{\sigma*}$ are
again left understood.  All the parameters appearing in the
superpotential Eq.~\eqref{eq:superpotential} and in the Lagrangian
Eq.~\eqref{eq:soft_terms} (and equivalently in the first two lines of
Eq.~\eqref{eq:mass_basis}) are in principle complex quantities.
However, superfield phase redefinition allows to remove several
complex phases. Here for simplicity, we will concentrate on soft
leptogenesis arising from a single sneutrino generation $i=1$ and in
what follows we will drop that index ($Y_\alpha\equiv Y_{1\alpha},\,
\widetilde{N}_{\pm}\equiv \widetilde{N}_{\pm 1}$, etc.).  
After superfield phase rotations, the relevant
Lagrangian terms restricted to $i=1$ are characterized by only two
independent physical phases:
\begin{eqnarray}
\label{eq:CPphase1}
\phi_{A}&=&{\rm arg}(A B^*),  \\
\phi_{g}&=&\frac{1}{2}{\rm arg}(B m_{2}^*), 
\label{eq:CPphase2}
\end{eqnarray}
which we choose to assign respectively to the slepton-Higgs-sneutrino
trilinear soft breaking terms, and to the gaugino coupling operators
respectively.  In what follows we will keep track of these
physical phases explicitly and, differently from the convention used in
Eqs.~\eqref{eq:superpotential},~\eqref{eq:soft_terms}
and~\eqref{eq:mass_basis}, we will leave understood (unless when
explicitly stated in the text) that all the other parameters
$Y_\alpha,B,\,m_{2},\,A$ etc. correspond to real and positive values.

\subsection{CP asymmetries in soft leptogenesis}
\label{sec:softCP}

Neglecting supersymmetry-breaking effects in the heavy sneutrino masses
and in the vertex, the total singlet sneutrino decay width is given by  
\begin{equation}
\Gamma_{\widetilde{N}_+}
=\Gamma_{\widetilde{N}_-}\equiv \Gamma
=\frac {M}{4\pi} {\displaystyle \sum_\alpha Y_{\alpha}^2}\equiv
\frac {m_{\rm eff}\, M^2}{4 \pi\, v_u^2}\,,  
\label{eq:gamma}
\end{equation}
where $v_u=v\, \sin\beta $ (with $v$=174 GeV) 
is the vacuum expectation value of the up-type Higgs
doublet and $m_{\rm eff}
\equiv \sum_\alpha Y^2_\alpha v_u^2/M$ is the rescaled decay width, that 
is related to the washout parameter $K$ as
$K=\Gamma_{\widetilde{N}}/H(M)=m_{\rm eff}/m_*$, where the equilibrium
mass is defined as $m_*=\sqrt{\frac{\pi g^*}{45}}\times\frac{8\pi^2
  v^2_u}{m_P}\sim 10^{-3}\,\,{\rm eV}$ with $g^*$ the total number
of relativistic degrees of freedom ($g^*=228.75$ in the MSSM).

There are three contributions to the CP asymmetry in
$\widetilde{N}_{\pm}$ decays into fermions  
$(\widetilde{H}_u,\,\ell_\alpha)$ 
and other three for decays into bosons
$(H_u,\,\widetilde{\ell}_\alpha)$~\cite{soft3,ourgaugino}.
Denoting the total CP asymmetries into the scalar and fermion channels
respectively with $s$ and $f$  subscripts, they are:
$\epsilon_{s,f}^{S}$ arising from self-energy diagrams induced
by the bilinear $B$ term; $\epsilon_{s,f}^{V}$ arising from
vertex diagrams induced by the gaugino masses;
$\epsilon_{s,f}^{I}$ arising from the interference of
  self-energy and vertex diagrams.  They can be written as: 
\begin{eqnarray}
\frac{\epsilon^S_{f}(T)}{\Delta_f(T)} &=& 
\frac{A}{M}\frac{4B\Gamma}{4B^2+\Gamma^2}
\left(1+\frac{\widetilde{M}^2}{M^2}-\frac{B^2}{2M^2}\right)\sin\phi_{A},
\label{eq:CP_asymresf} 
\\
\frac{\epsilon^S_{s}(T)}{\Delta_s(T)}
 &=& -\frac{A}{M}\frac{4B\Gamma}{4B^2+\Gamma^2}
\left(1-\frac{A^2}{M^2}\right)\sin\phi_{A},
\label{eq:CP_asymress}  
\\
\frac{\epsilon_{f}^{V}\left(T\right)}{\Delta_f(T)}
= -\frac{\epsilon_{s}^{V}\left(T\right)}{\Delta_s(T)}
 &=&  \frac{3\alpha_{2}}{4}
\frac{A}{M} \frac{m_{2}}{M}
\left(\ln\frac{m_{2}^{2}}{m_{2}^{2}+M^{2}}\right) 
\left\{
\sin(\phi_{A}+2\phi_{g})
-\frac{B}{A}\sin\left(2\phi_{g}\right)\right\}\quad  
\label{eq:CP_asymver}  
\\
\frac{\epsilon_{f}^{I}\left(T\right)}{\Delta_f(T)}
= -\frac{\epsilon_{s}^{I}\left(T\right)}{\Delta_{s}\left(T\right)}
&=&- \frac{3\alpha_{2}}{2}
\frac{A}{M}\frac{m_{2}}{M}
\frac{\Gamma^{2}}{4B^{2}+\Gamma^{2}}
\left(\ln\frac{m_{2}^{2}}{m_{2}^{2}+M^{2}}\right)
\sin\phi_{A} \cos\left(2\phi_{g}\right), 
\label{eq:CP_asymint} 
\end{eqnarray}
where $\alpha_2=\frac{g_2^2}{4\pi}$.  To take into account the effects
of lepton flavours, we need to define instead
of~\Eqns{eq:CP_asymresf}{eq:CP_asymint} the corresponding
asymmetries for $\widetilde{N}_\pm$ decays into fermions and scalars
of a specific lepton flavour $\ell_\alpha$ and
$\widetilde{\ell}_\alpha$.  The assumption of universality of the soft
terms implies that the flavour CP asymmetries are simply related to
the total CP asymmetries through the corresponding decay branching
fractions, denoted  by $P_\alpha$:
\begin{equation}
\epsilon^{S,V,I}_{\alpha\,(s,f)}= P_\alpha\; \epsilon^{S,V,I}_{(s,f)}\;,
\label{eq:flavorCP}
\end{equation}
where, in terms of the Yukawa couplings, the branching fractions can
be written as:
\begin{equation}
\label{eq:Palpha}
P_\alpha \equiv  \frac{ Y_{\alpha}^2}{\displaystyle 
\sum_\beta Y_{\beta}^2},\qquad \qquad 
\sum_\alpha P_\alpha=1.  
\end{equation}
It is worth recalling that when the condition of universality of the
soft terms is dropped the simple relation~\Eqn{eq:flavorCP} does not
hold anymore.  The expressions for the flavoured CP asymmetries in the
general case of non-universal soft terms can be found
in~\cite{oursoftnu}.  The $\Delta_{s,f}(T)$ terms
in~\Eqns{eq:CP_asymresf}{eq:CP_asymint} denote the scalar and fermion
thermal factors that are related to thermal phase-space, Bose
enhancement and Fermi blocking. They are the same for $\widetilde
N_\pm$ and are normalized so that their zero temperature limit is
$\Delta_{s,f}(T\!=\!0)=\frac{1}{2}$. Their explicit expression is
given in Appendix~\ref{Appendix-A}.  Note that in
writing~\Eqns{eq:CP_asymresf}{eq:CP_asymint} we have implicitly
assumed that the thermal factors are flavour independent. This is
indeed an excellent approximation as long as zero temperature slepton
masses and small Yukawa couplings are neglected.

\Eqns{eq:CP_asymresf}{eq:CP_asymint} are approximate expressions with
only the leading terms in $\frac{\Lambda_{susy}}{M}$ included.  In
Eqs.~\eqref{eq:CP_asymresf} and ~\eqref{eq:CP_asymress} however, we
have included also $\mathcal{O}(\frac{\Lambda_{susy}^2}{M^2})$
corrections, since when the resonant condition $\Gamma\sim 2 B$ is
satisfied $\epsilon^S$ is the dominant CP asymmetry.  Note that with
these corrections included the self-energy CP asymmetries for fermions
and scalars do not cancel anymore in the $T\to 0$ limit.  In contrast,
as we will argue in the next section, for the vertex and interference
asymmetries $\epsilon^{V,I}$ at one loop the cancellation between
scalar and fermion contributions is exact.  Neglecting for simplicity
the higher order terms in Eqs.~\eqref{eq:CP_asymresf} and
~\eqref{eq:CP_asymress}, the total CP asymmetry summed over scalars
and fermions final states can be written as
\begin{equation}
\epsilon_\alpha (T)\equiv P_\alpha\,  \bar\epsilon \cdot 
\big[\Delta_s(T)-\Delta_f(T)\big] 
\label{eq:eptot}
\end{equation}
where  $\bar\epsilon$ is independent of $T$ and 
\begin{equation}
\big[\Delta_s-\Delta_f\big]\ \  \stackrel{T\to 0}{\longrightarrow}\ \ 0\;. 
\label{eq:Tto0}
\end{equation}

\subsection{Vanishing of the CP asymmetry in decay}
\label{sec:vertexCP}
The new sources of direct CP violation from vertex corrections
involving the gauginos were first introduced in soft leptogenesis
in~\cite{soft3}. In the same paper it was also stated that the new
contributions do not require thermal effects to produce a sizable
lepton asymmetry in the plasma.  Gaugino contributions to soft
leptogenesis were reconsidered in~\cite{ourgaugino} where it was
instead found that the zero temperature cancellation between the CP
asymmetries for decays into scalars and into fermions holds also when
vertex corrections are included.  This issue is of some interest,
because if thermal corrections are necessary for soft leptogenesis to
work, then non-thermal scenarios, like the ones in which sneutrinos
are produced by inflaton decays and the thermal bath remains at a
temperature $T \ll M$ during the following leptogenesis epoch, would
be completely excluded.  Here we present a simple but general argument
proving that the direct leptonic CP violation in sneutrinos decays
vanishes at one loop, due to an exact cancellation between the scalar
and fermion contributions, in agreement with the explicit calculation in 
\cite{ourgaugino}.  We should also clarify that while above
the limit in~\Eqn{eq:NSE} soft leptogenesis can be successful even
when the $T\to 0$ limit for the decay CP asymmetries is taken, this
happens because of other effects which are linked to the washout
processes, and thus do require a thermal bath. Therefore, the fact
that soft leptogenesis cannot work in non-thermal scenarios is always
true.

\begin{figure}[t!]
\includegraphics[width=\textwidth]{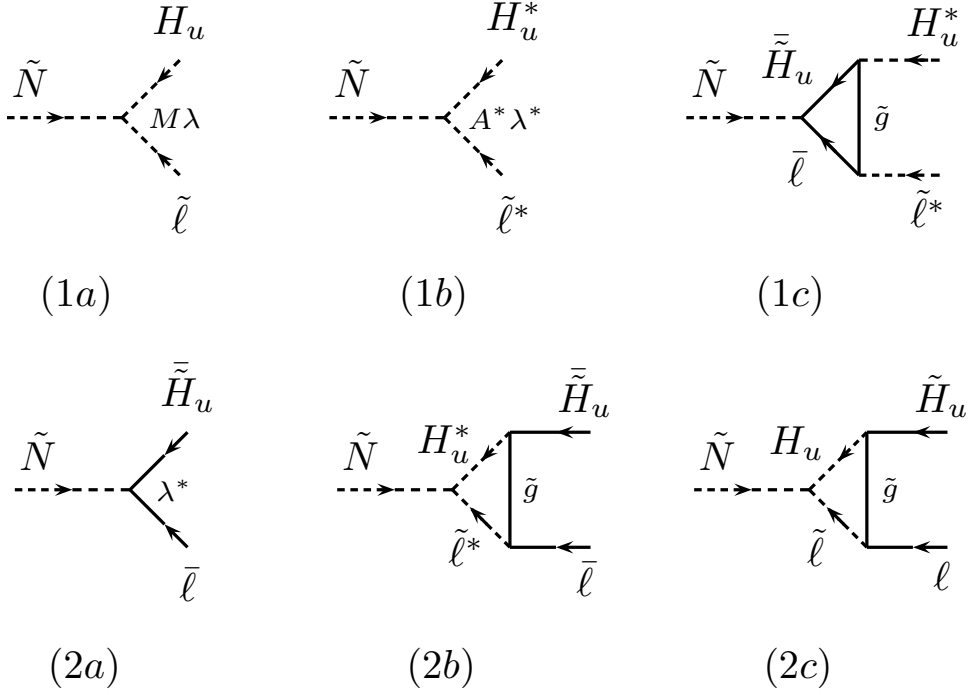}
\caption[]{ 
Soft leptogenesis  diagrams for sneutrino decays into
scalars $(1a),\,(1b),\,(1c)$ and into fermions 
$(2a),\,(2b),\,(2c).$}
\label{fig:1}
\end{figure} 


Let us take for simplicity $\Phi=0$ in~\Eqn{eq:mass_eigenstates} (this
amounts to assign the phases $\phi_A$ and $\phi_g$ in
\Eqn{eq:CPphase1} and \Eqn{eq:CPphase2} respectively to $A$ and
$m_2$), and let us introduce for the various amplitudes the shorthand
notation $A^{\pm}_\ell \equiv A(\widetilde{N}_\pm\to \ell \widetilde
H_u)$, $A^{\widetilde{N}\, (\widetilde{N}^*)}_\ell \equiv
A\left(\widetilde{N}\,(\widetilde{N}^*) \to \ell \widetilde
  H_u\right)$ with similar expressions for the other final
states. From~\Eqn{eq:mass_eigenstates} we can write
\begin{eqnarray}
  \label{eq:Apmell}
 2\, \left|A^\pm_\ell\right|^2 &=& 
\left|A^{\widetilde{N}}_\ell\right|^2+
\left|A^{\widetilde{N}^*}_\ell\right|^2 
\pm  2\, \mathrm{Re}
\left(A^{\widetilde{N}}_\ell
\cdot 
A^{\widetilde{N}}_{\bar\ell}\right)
\\
  \label{eq:Apmbell}
 2\, \left|A^\pm_{\bar \ell}\right|^2 &=& 
\left|A^{\widetilde{N}}_{\bar \ell}\right|^2+
\left|A^{\widetilde{N}^*}_{\bar \ell}\right|^2 
\pm  2\, \mathrm{Re}
\left(A^{\widetilde{N}}_{\bar \ell}
\cdot 
A^{\widetilde{N}}_{\ell}\right)\,, 
\end{eqnarray}
where the complex conjugate amplitudes in the last terms of both these
equations have been rewritten as follows:
$(A^{\widetilde{N}^*}_\ell)^*= A_{\widetilde{N}^*}^\ell=
A^{\widetilde{N}}_{\bar\ell}$ and $(A^{\widetilde{N}^*}_{\bar
  \ell})^*= A_{\widetilde{N}^*}^{\bar \ell}= A^{\widetilde{N}}_{\ell}$
by using CPT in the second step.  The direct CP asymmetry for
$\widetilde{N}_\pm$ decays into fermions is given by the difference
between \Eqn{eq:Apmell} and \Eqn{eq:Apmbell}:
\begin{equation}
  \label{eq:Af}
  2\,\left(
\left|A^\pm_{\ell}\right|^2-
\left|A^\pm_{\bar \ell}\right|^2\right)=
\left(\left|A^{\widetilde{N}^*}_\ell\right|^2-
\left|A^{\widetilde{N}^*}_{\bar \ell}\right|^2\right)
+
\left(\left|A^{\widetilde{N}}_\ell\right|^2-
\left|A^{\widetilde{N}}_{\bar \ell}\right|^2\right)\,.
\end{equation}
With the replacements $\ell \to \tilde\ell$ and 
$\bar\ell\to\tilde\ell^*$, a completely equivalent expression
holds also for the decays into scalars. 

The tree level and one loop diagrams for the various decay amplitudes
into scalars and fermions are given in Figure~\ref{fig:1}.  We note at
this point that 
$A^{\widetilde N}_{\widetilde
  \ell}$ has no one-loop amplitude to interfere with 
(see diagram $(1\,a)$) 
and thus, up to
one-loop, the full amplitude coincides with the tree level result, and
is CP conserving.  $A^{\widetilde N}_{\ell}$ is a pure
one-loop amplitude (see diagram $(2\,c)$) and therefore  is also CP
conserving. This implies:
\begin{eqnarray}
  \label{eq:tree}
\left| A^{\widetilde N}_{\widetilde \ell}\right|^2 &=&  
 \left|  A^{\widetilde N^*}_{\widetilde \ell^*}\right|^2\, \\ 
  \label{eq:loop}
\left|A^{\widetilde N}_{\ell}\right|^2  &=&
  \left| A^{\widetilde N^*}_{\bar \ell}\right|^2. 
\end{eqnarray}
We can thus change simultaneously the signs of $\left|A^{\widetilde
    N}_{\ell}\right|^2$ and $\left| A^{\widetilde N^*}_{\bar
    \ell}\right|^2$ in~\Eqn{eq:Af} without affecting the equality, and
the same we can do in the analogous equation for the 
scalars. This gives:
\begin{eqnarray}
  \label{eq:Af1}
  2\,\left(
\left|A^\pm_{\ell}\right|^2-
\left|A^\pm_{\bar \ell}\right|^2\right)&=&
\left(\left|A^{\widetilde{N}^*}_\ell\right|^2+
\left|A^{\widetilde{N}^*}_{\bar \ell}\right|^2\right)
-
\left(\left|A^{\widetilde{N}}_\ell\right|^2+
\left|A^{\widetilde{N}}_{\bar \ell}\right|^2\right)\,, \\
  \label{eq:As1}
  2\,\left(
\left|A^\pm_{\tilde \ell}\right|^2-
\left|A^\pm_{\tilde \ell^*}\right|^2\right)&=&
\left(\left|A^{\widetilde{N}^*}_{\tilde\ell}\right|^2+
\left|A^{\widetilde{N}^*}_{\tilde \ell^*}\right|^2\right)
-
\left(\left|A^{\widetilde{N}}_{\tilde\ell}\right|^2+
\left|A^{\widetilde{N}}_{\tilde \ell^*}\right|^2\right)\,.
\end{eqnarray}
Using CPT $A^{\widetilde{N}^*}_{\rm all} = A_{\widetilde{N}}^{\rm
  all}$ and unitarity $\left|A_{\widetilde{N}}^{\rm all}\right|^2 =
\left|A^{\widetilde{N}}_{\rm all}\right|^2 $ we can readily see that
the sum of these two equations vanishes.  We have thus proved that for
$\widetilde{N}_+$ and $\widetilde{N}_-$ independently, at one loop
there is an exact cancellation between the scalars and fermions final
state contributions, and thus at $T=0$ the direct decay CP
asymmetries vanish.

\section{Soft leptogenesis above the superequilibration temperature}
\label{sec:NSE}

We now discuss the early Universe effective theory appropriate for
studying soft leptogenesis in the regime in which superequilibrating
reactions like $\widetilde\ell\, \widetilde\ell \to \ell \ell$, that
are induced by gaugino masses, and higgsino mixing transitions, that
are induced by the supersymmetric $\muh H_u H_d$ term, do not occur.
For simplicity, we assume equal masses for all the gauginos
$m_1=m_2=m_3=m_{\tilde g}$ and that the supersymmetric higgsino mixing
term also has approximately the same value: $\muh \simeq m_{\tilde g}
= \Lambda_{susy}$.  The regime we are interested in is defined by the
condition given in \Eqn{eq:NSE}, that is the lower limit on the
relevant temperatures is:
\begin{equation}
T \gtrsim 5\cdot 10^7 
\left(\frac{\Lambda_{susy}}{500\,{\rm GeV}}\right)^{2/3}\;{\rm GeV}.
\label{eq:Tgmu}
\end{equation}

\subsection{Anomalous and non anomalous symmetries} 
\label{sec:symmetries}

The supersymmetric effective theory appropriate to study particle
physics processes in the early Universe when the thermal bath
temperature satisfies the condition~\Eqn{eq:Tgmu} is obtained by
setting $m_{\tilde g}, \muh \to 0$~\cite{Ibanez:1992aj}.  
In this limit the theory gains two new $U(1)$ symmetries: $\muh\to 0$
yields a global symmetry of the Peccei-Quinn (PQ) type, and by setting
also $m_{\tilde g}\to 0$ one additional global $R$-symmetry arises.

\begin{table}[t!!]
 \begin{center} 
\begin{tabular}{|cc|c|c|c|c|c|c|c|c|c|c|c|c|}
 \hline
&
&\quad  $\tilde g\ $
&\quad $Q\ $
&\quad$u^c \ $
&\quad $d^c \ $
&\quad $\ell \ $
&\quad $e^c \ $
&\quad $\tilde H_d\phantom{\Big|}$
&\quad$\tilde H_u\ $
&\quad$N^c\  $
\\
\hline
\multicolumn{2}{|c|}{$B\phantom{\Big|}$} &$ 0$&$ \frac{1}{3}$&$-\frac{1}{3}$&$-\frac{1}{3}$&$ 0$&$ 0$&$ 0$&$ 0$& 0 \\ \hline 
\multicolumn{2}{|c|}{$L\phantom{\Big|}$} &$ 0$&$ 0$          &$ 0$          &$ 0$          &$ 1$&$-1$&$ 0$&$ 0$& 0 \\ \hline 
\multicolumn{2}{|c|}{$PQ\phantom{\Big|}$}&$ 0$&$ 0$          &$-2$          &$ 1$          &$-1$&$ 2$&$-1$&$ 2$& 0 \\ \hline 
\multirow{2}{*}{$R$}&$\phantom{\!\!\!\!\Big|}f$ &$ 1$&$-1$   &$-3$          &$ 1$          &$-1$&$ 1$&$-1$&$3$&$-1$ \\ \cline{3-11}
 &$\phantom{\!\!\!\!\Big|}b$             &$ 2$&$ 0$          &$-2$          &$ 2$          &$ 0$&$ 2$&$ 0$&$ 4$& 0 \\ \hline
\end{tabular}
\caption{$B$, $L$, $PQ$ and $R$ charges for the particle supermultiplets 
  that are labeled in the top row by their L-handed fermion
  component.  Note that we use chemical potentials for the R-handed
  $SU(2)$ singlet fields $u,\,d,\,e$ that have opposite charges with
  respect to the ones for $u^c,\,d^c,\,e^c$ given in the
  table. The $R$ charges for bosons are determined by $R(b)=R(f)+1$.}
\label{tab:1}
\end{center}
\end{table}

The charges of the various states under $R$ and $PQ$, together with
the values of the other two global symmetries $B$ and $L$ are given in
Table~\ref{tab:1}.  Like $L$, also $R$ and $PQ$ are not symmetries of
the seesaw superpotential terms $MN^cN^c+ \lambda N^c\ell H_u$, since
it is not possible to find any charge assignment that would leave both
terms invariant.  In Table~\ref{tab:1} we have fixed the charges of
the heavy $N^c$ supermultiplets in such a way that sneutrinos do not
carry any charge.\footnote{This differs from the assignments adopted
  in Ref.~\cite{ournse}.} This has the advantage of ensuring that all
the sneutrino bilinear terms, corresponding to the mass parameters
$M,\,\widetilde{M},\,B$, are invariant, and thus sneutrino mixing does
not break any symmetry. However, since $R(N^cN^c)=0$, it follows that
the mass term for the heavy Majorana neutrino breaks $R$ by two
units.\footnote{Under $R$-symmetry the superspace Grassmann parameter
  transform as $\theta \to e^{i\alpha}\theta$ . Invariance of $\int
  d\theta\, \theta =1$ then requires $R(d\theta)=-1$. Then the chiral
  superspace integral of the superpotential $\int d\theta^2\, W $ is
  invariant if $R(W)=2$. By expanding a chiral supermultiplet in
  powers of $\theta$ it follows that the supermultiplet $R$ charge
  equals the charge of the bosonic scalar component $R(b)=R(f)+1$, and
  thus for the fermion bilinear term
  $R\left(\overline{N^c_R}N^c_L\right)=-2$.}

All the four global symmetries $B$, $L$, $PQ$ and $R$ have mixed gauge
anomalies with $SU(2)$, and $R$ and $PQ$ have also mixed gauge
anomalies with $SU(3)$.  Two linear combinations of $R$ and $PQ$,
having respectively only $SU(2)$ and $SU(3)$ mixed anomalies, have been
identified in Ref.~\cite{Ibanez:1992aj}.  They are:~\footnote{ With
  respect to Ref.~\cite{Ibanez:1992aj}, for definiteness we restrict
  ourselves to the case of three generations $N_g=3$ and one pair of
  Higgs doublets $N_h=1$, and we also normalize $R_{2,3}$ in such a
  way that $R_{2,3}(b) =R_{2,3}(f)+1$.}
\bea
\label{eq:R2}
R_2 &=& R-2\,PQ \\
\label{eq:R3}
R_3 &=&R-3\,PQ\,. 
\eea
The values of $R_{2,3}$ for the different states are given in
Table~\ref{tab:2}.  The authors of Ref.~\cite{Ibanez:1992aj} have also
constructed the effective multi-fermions operators generated by the
mixed anomalies: \bea
\label{eq:tO-EW}
\tilde O_{EW} &=&
\Pi_\alpha \left(QQQ\ell_\alpha\right)\; \tilde H_u\tilde H_d\;\tilde W^4\,,\\
\label{eq:tO-QCD}
\tilde O_{QCD} &=& \Pi_i \left(QQu^c d^c\right)_{i}\; \tilde g^6 \,.
\eea
Given that we have three charges $R_2$, $B$ and $L$ with mixed $SU(2)$
anomalies, it is then possible to define two anomaly free
combinations.  The most convenient are $B-L$ and
\be
\label{eq:RB}
 {R_B}=\frac{2}{3}B+R_2,  
\ee
whose values are also given in Table~\ref{tab:2}.  For the problem at
hand, $R_B$ is more convenient than the charge ${\cal
  R}=\frac{5}{3}B-L+R_2=R_B-(B-L)$ that was introduced
in~\cite{ournse}. This is because ${\cal R}$ is not conserved in
sneutrino decays that are induced by sneutrino-related soft terms, and
so it does not correspond any more to a global neutrality
condition~\cite{ournse}. On the other hand, the fact that $R_B$ does
not contain any $B-L$ fragment, ensures that 
it will not enter in the final computation of the baryon asymmetry that 
will only depend on $B-L$. The fact that $R_B$ is independent of $L$ renders
also easier writing a BE for its evolution.

The $R_B$ values in Table~\ref{tab:2} imply that the superpotential
term $N^c\,\ell H_u$ has charge $R_B=2$ and thus is invariant. It
follows that sneutrinos decays into fermions conserve $R_B$.  In
contrast, the soft $A$ term in~\Eqn{eq:soft_terms} responsible for
sneutrinos decays into scalars violates $R_B$ by 2 units, more
precisely $\widetilde{N}_\pm \to H_u\tilde\ell$ has $\Delta R_B=+2$
while $\widetilde{N}_\pm \to H_u^*\tilde\ell^*$ has $\Delta R_B=-2$.
As regards the heavy neutrinos, their mass term violates $R_B$ by two
units. Note that this is precisely like the case when one chooses
to assign a lepton number $-1$ to the singlet neutrinos
$N$. Accordingly, the decays of the heavy Majorana neutrino violate
$R_B$ by one unit: $N\to \ell H_u,\, \tilde\ell\tilde H_u$ have
$\Delta R_B=+1$ and the decays to the CP conjugate states have $\Delta
R_B=-1$.  Since all $R_B$ violating reactions have, by assumption,
rates that are comparable to the Universe expansion rate, the
evolution of this charge must then be tracked by means of a specific
BE.

 \begin{table}[t!]
\begin{center}
\begin{tabular}{|cc|c|c|c|c|c|c|c|c|c|c|c|c|}
 \hline
&
&\quad  $\tilde g\ $
&\quad $Q\ $
&\quad$u^c \ $
&\quad $d^c \ $
&\quad $\ell \ $
&\quad $e^c \ $
&\quad $\tilde H_d\phantom{\Big|}$
&\quad$\tilde H_u\ $
&\quad$N^c\  $
\\
\hline
\multirow{2}{*}{$R_2$} &$\phantom{\!\!\!\!\Big|}f$     
&$ 1$&$-1$&$ 1$&$-1$&$ 1$&$ -3$&$1$&$-1$& $-1$\\ \cline{3-11}
                     &$\phantom{\!\!\!\!\Big|}b$       
&$ 2$&$ 0$&$ 2$&$ 0$&$ 2$&$-2$&$ 2$&$ 0$& 0 \\ \hline
\multirow{2}{*}{$R_3$} &$\phantom{\!\!\!\!\Big|}f$     
&$ 1$&$-1$&$ 3$&$-2$&$ 2$&$ -5$&$2$&$-3$& $-1$\\ \cline{3-11}
                     &$\phantom{\!\!\!\!\Big|}b$       
&$ 2$&$ 0$&$ 4$&$ -1$&$ 3$&$-4$&$ 3$&$-2$& 0 \\ \hline
\hline
\multirow{2}{*}{$R_B$}&$\phantom{\!\!\!\!\Big|}f$ 
&$ 1$&$-\frac{7}{9}$&$\frac{7}{9}$&$-\frac{11}{9}$&$1$&$ -3$&$
1$&$-1$& $-1$ \\
 \cline{3-11}   &$\phantom{\!\!\!\!\Big|}b$       
&$ 2$&$\frac{2}{9}$&$\frac{16}{9}$&$-\frac{2}{9}$&$ 2$&$-2$&$ 2$&$ 0$& 0 \\ 
\hline
\end{tabular}
\caption{Charges for the fermionic and bosonic components of the SUSY
  multiplets under the $R$-symmetries defined in Eqs.\protect
  \eqref{eq:R2}, \protect \eqref{eq:R3} and \protect \eqref{eq:RB}.
  Supermultiplets are labeled in the top row by their L-handed fermion
  component.  We use chemical potentials for the R-handed $SU(2)$
  singlet fields $u,\,d,\,e$ that have opposite charges with respect
  to the ones for $u^c,\,d^c,\,e^c$ given in the table.}
\label{tab:2}
\end{center}
\end{table}

At temperatures satisfying the condition~\Eqn{eq:Tgmu} there is at
least one other anomalous global symmetry, that we will denote by
$\chi$. It corresponds to $U(1)$ phase rotations of the $u^c$ chiral
multiplet that, for its fermionic component, can be readily identified
with chiral symmetry for the right-handed up-quark. In fact, above
$T\sim 2\times 10^{6}\,$GeV, reactions mediated by $h_u$ do not occur
and the condition $h_u\to 0$ must be imposed, resulting in a new
anomalous `chiral' symmetry.  In the $SU(3)$ sector we then have two
anomalous symmetries $R_3$ and $\chi$, and one anomaly free
combination can be constructed. Assigning to the $L$-handed $u^c_L$
supermultiplet a chiral charge $\chi=-1$ this combination has the
form~\cite{ournse}
\begin{equation}
\label{eq:chiralup}
R_\chi= \raise 2pt \hbox{$\chi$}_{u^c_L}+\kappa_{u^c_L}\,R_3, 
\end{equation}
where $\kappa_{u^c_L}=1/3$. When the additional condition $h_d\to 0$
is imposed, a chiral symmetry arises also for the $d^c$
supermultiplet.  A second anomaly free $R_\chi$ symmetry can then be
defined in a way completely analogous to \Eqn{eq:chiralup}, with
$\kappa_{d^c_L}=\kappa_{u^c_L}=1/3$~\cite{ournse}.  As regards
perturbative violations of $R_\chi$, this charge inherits the same
violations $R_3$ suffers.  The soft $A$ term in~\Eqn{eq:soft_terms}
violates $R_3$ by one unit, and so do sneutrinos decays into scalars.
Moreover, since $N^c\,\ell H_u$ has an overall charge $R_3=1$, a
violation by one unit occurs also for sneutrinos decays into
fermions. Correspondingly, we have $\Delta R_3=+1$ for the decays
$\tilde N,\,\tilde N^* \to H_u\tilde\ell,\; \overline{\widetilde{H}}_u
\overline\ell $ and $\Delta R_3=-1$ for $\tilde N,\,\tilde N^* \to
\widetilde{H}_u \ell,\; H_u^*\tilde\ell^*$. Of course, similarly to
$R_B$, also the evolution of $R_\chi$ needs to be tracked by means of
a BE.


\subsection{Chemical equilibrium conditions and conservation laws}

Because of the network of fast particle reactions occurring in the
thermal bath, asymmetries generated in sneutrino decays spread around
among the various particle species, and this can affect directly or
indirectly leptogenesis processes.  In principle there is one
asymmetry for each particle degree of freedom. There are however
several conditions and constraints that reduce the number of
independent asymmetries to a few. The three types of reactions that
have been classified in the introduction give rise to three different
types of constraints and conditions, that need to be formulated in
their own appropriate way:

\begin{itemize}
\item[(i)] Constraints imposed by reactions whose rates are much  
faster than the Universe expansion have to be formulated in
terms of chemical equilibrium conditions for the chemical potentials of
incoming $\mu_I$ and  final state particles $\mu_F$:
\begin{equation}
  \label{eq:muimuf}
  \sum_I \mu_I = \sum_F \mu_F.
\end{equation}

\item[(ii)] Conservation laws that  arise when all the reactions 
that violate some specific charge are much slower than the 
the Universe expansion have to be formulated in terms of particle 
number densities $\Delta n = n - \bar n$ and, for a generic charge
$Q$, read: 
\begin{equation}
  \label{eq:Qtot}
  Q=\sum_i   Q_i \Delta n_i = {\rm const}, 
\end{equation}
where $Q_i$ is the charge of the $i$-particle species. 
We will always assume as initial conditions for leptogenesis 
that all particle asymmetries vanish, and thus we will put the 
constant value of \Eqn{eq:Qtot} equal to zero.

\item[(iii)] Reactions with rates comparable with the Universe
  expansion  have to be treated by means of appropriate dynamical
  equations. In this case, in order to reabsorb the dilution effects due to the
  Universe expansion, it is convenient to introduce as basic variables
  the number densities of particles per degree of freedom $g$ normalized 
to the entropy density $s$: 
\begin{equation}
  \label{eq:Y}
  Y_{\Delta_i} = \frac{1}{g_i}\frac{\Delta n_i}{s}\,. 
\end{equation}
\end{itemize}

Clearly, $\mu_i$, $\Delta n_i$ and $Y_{\Delta_i}$ are all related to
particles asymmetries. In particular, the number density asymmetries
of particles for which a chemical potential can be defined are
directly related with this chemical potential. For both bosons ($b$)
and fermions ($f$) this relation acquires a particularly simple form
in the relativistic limit $m_{b,f}\ll T$, and at first order in
$\mu_{b,f}/T\ll 1$:
\be
\label{eq:Dnmu}
\Delta n_{b}=\frac{g_b}{3}
  T^2\mu_b, \qquad\quad
\Delta n_{f}=\frac{g_f}{6}T^2\mu_f\,.
\ee
While we will always express the various constraints using the most
appropriate quantities, eventually to solve for the large set of
conditions in a closed form we will need to use a single set of
variables. We will take this to be the set
$\left\{Y_{\Delta_i}\right\}$, and will leave understood that our
solutions to the constraining conditions are obtained after expressing
$\mu_i$ and $\Delta n_i$ in terms of this set, through \Eqn{eq:Dnmu}
and \Eqn{eq:Y}.

In the following we denote the chemical potentials with the same
notation that labels the corresponding field: $\phi \equiv \mu_{\phi}$
and, for definiteness, we fix the relevant values of the temperature
around $T\sim 10^{8}\,$GeV.  The set of conditions that constrain the
particle abundances at this temperatures are listed below, more
details about the various constraints can be found in~\cite{ournse}.


\begin{enumerate}

\item At $T\gg M_W$, gauge fields have vanishing chemical potential
  $W=B=g=0$~\cite{ha90}. This also implies that particles belonging to
  the same $SU(2)$ or $SU(3)$ multiplets have the same chemical
  potential. For example
  $\phi(I_3=+\frac{1}{2})=\phi(I_3=-\frac{1}{2})$ for a field $\phi$
  that is a doublet of weak isospin $\vec I$, and similarly for color.

\item Denoting by $\tilde W_R$, $\tilde B_R$ and $\tilde g_R$ the
  right-handed winos, binos and gluinos chemical potentials, and by
  $\ell,\,Q$ ($\tilde\ell,\,\tilde Q$) the chemical potentials of the
  (s)lepton and (s)quarks left-handed doublets, the following
  reactions: $\tilde Q +\tilde g_R \to Q$,\ $\tilde Q +\tilde W_R \to
  Q$,\ $\tilde \ell +\tilde W_R \to \ell $,\ $\tilde \ell +\tilde B_R
  \to \ell $,\ 
  imply that all gauginos have the same chemical potential:
\begin{equation}
\label{eq:g}
-\tilde g = Q-\tilde Q=
-\tilde W= \ell-\tilde \ell=-\tilde B,
\end{equation}
where we have introduced $\tilde W$, $\tilde B$ and $\tilde g$
to denote the chemical potential of the {\it left-handed} gauginos.
It follows that the chemical potentials of the SM particles are related 
to the chemical potential of their respective superpartners as 
\
\begin{eqnarray}
  \label{eq:tQtell}
   \tilde{Q},\tilde \ell &=&    Q,\ell+  \tilde g \\
  \label{eq:HuHd}
   H_{u,d} &=&   \tilde H_{u,d}+  \tilde g \\
  \label{eq:tutdte}
   \tilde u,\tilde d,\tilde e  &=&   u,d,e-  \tilde g. 
\end{eqnarray}
The last relation, in which $u,d,e\equiv u_R,d_R,e_R$ denote the
$R$-handed $SU(2)$ singlets, follows e.g. from $ \tilde u^c_L= u^c_L+
\tilde g$ for the corresponding $L$-handed fields, together with
$u^c_L=-u_R$, and from the analogous relation for the $SU(2)$ singlet
squarks.  

\end{enumerate}

\smallskip

Eqs.\eqref{eq:tQtell}--\eqref{eq:tutdte} together with the vanishing
of the chemical potentials of the gauge fields and the equality of the
chemical potentials for all the gauginos, implies that we are left
with 18 chemical potentials (or number density asymmetries) that we
chose to be the ones of the fermionic states. They are 15 for the SM
quarks and leptons, 2 for the up-type and down-type higgsinos, and 1
for the gauginos. These 18 quantities are further constrained by
additional conditions.

\smallskip

\begin{enumerate}

\item[3.] Before EW symmetry breaking hypercharge is an exactly conserved
  quantity.  Therefore for the total hypercharge of the Universe we
  have 
  \begin{equation}
    \label{eq:Ytot}
    {y}_{\,\rm tot} = \sum_{b} \Delta n_b\, y_b +   \sum_{f} \Delta
    n_f\,y_f =0, 
  \end{equation}
  where $y_{b,f}$ denotes the hypercharge of the $b$-bosons or
  $f$-fermions.  It is useful to rewrite explicitly 
this condition in terms of the rescaled density asymmetries per
  degree of freedom $\left\{Y_{\Delta_i}\right\}$ defined in \Eqn{eq:Y}:
 \begin{equation}
    \label{eq:YtotY}
\sum_i\left(Y_{\Delta Q_i}+2Y_{\Delta u_i}-Y_{\Delta d_i}\right)
-\sum_\alpha\left(Y_{\Delta \ell_\alpha}+Y_{\Delta e_\alpha}\right)+
Y_{\Delta \tilde{H}_u}-Y_{\Delta \tilde{H}_d}= 0. 
\end{equation}

\item[4.] Chemical equilibrium for reactions that are mediated by the
  leptons and quarks Yukawa couplings give:
\bea 
\label{eq:leptons}
\ell_\alpha - e_\alpha + \tilde H_d + \tilde g &=&0, \qquad
(\alpha=e,\,\mu,\,\tau), \\  
\label{eq:downquarks}
Q_i - d_i + \tilde H_d + \tilde g &=&0, \qquad
(d_i=d,\,s,\,b)\,, \\  
\label{eq:upquarks}
Q_i - u_i + \tilde H_u + \tilde g &=&0, \qquad
(u_i=c,\,t). 
\eea
At $T \sim 10^{8}\,$GeV, Yukawa
equilibrium for the up quark is never realized.  For $\alpha=e$ and
for the $d$-quark Yukawa equilibrium holds as long as $ T\lsim
10^5(1+\tan^2\beta)\,{\rm GeV}$~\cite{eR-equilibrium} and $ T\lsim
4\cdot 10^6(1+\tan^2\beta)\,{\rm GeV}$ respectively.  Then, for $T
\sim 10^{8}\,$GeV both condition hold only if $\tan\beta \gsim 35$,
while they both do not hold if $\tan\beta \lsim 5$. As we will discuss
below, in the latter case the Yukawa equilibrium conditions get
replaced by other two conditions, and thus the overall number of
constraints does not change.  Below we present results for the large
and small $\tan\beta$ cases, and since they do not differ much, we
omit the corresponding results for the intermediate case $5\lsim \tan
\beta\lsim 35$.  \\
Besides the previous Yukawa equilibrium conditions, quark
intergenerational mixing guarantees that below $T\lsim 10^{11}\,$GeV
the three quark doublets have the same chemical
potential:
  \begin{equation}
    \label{eq:Q}
    Q\equiv Q_3=Q_2=Q_1. 
  \end{equation}

\item[5.] Finally, reactions induced by the QCD and EW sphaleron
  multi-fermion operators~\Eqn{eq:tO-EW} and~\Eqn{eq:tO-QCD}
  imply~\cite{Ibanez:1992aj}
\bea
\label{eq:tEWmu}
&&3\sum_i Q_i+\sum_\alpha\ell_\alpha
+\tilde H_u+\tilde H_d+4\,\tilde g=0,\\
\label{eq:tQCDmu}
&&2\sum_i Q_i-\sum_i\left(u_i+d_i\right)+6\,\tilde g=0.
\eea

\end{enumerate}

Counting the number of additional conditions listed in items 3 to 5,
we have 1 from global hypercharge neutrality, 8 from Yukawa
equilibrium plus 2 due to quark intergenerational mixing, and 2 from
the EW and QCD sphaleron equilibrium. This adds to a total of 13
constraints for the initial 18 variables, meaning that 5 quantities
must be determined from dynamical evolution equations.  These
quantities can be chosen, for example, as the density-asymmetries of
the three lepton flavours $Y_{\Delta\ell_\alpha}$, of the up-type
higgsinos $Y_{\Delta \tilde H_u}$ and of the gauginos $Y_{\Delta\tilde
  g}$, where the last one allows to relate the previous four
quantities to the corresponding densities asymmetries of their
superpartners.  This choice would be a natural one since these are the
density asymmetries that `weight' the various interactions entering
the BE for soft leptogenesis.  However, the EW and QCD sphalerons
reactions~\Eqn{eq:tO-EW} and \Eqn{eq:tO-QCD} imply fast changes of
these asymmetries. A much more convenient choice is instead that of
using appropriate linear combinations of the various asymmetries
corresponding to anomaly free and quasi-conserved charges, where with
`quasi-conserved' we refer to charges that are not conserved only by
the `slow' sneutrino-related reactions.  These quantities can be
identified with the three flavoured leptonic charges
$\Delta_\alpha=B/3-L_\alpha$ and with the two $R_B$ and
$R_\chi$ charges discussed in the previous section. In terms of the rescaled
density asymmetries per degree of freedom they read:

\begin{eqnarray} 
\label{eq:YDeltaAlpha}
Y_{\Delta_\alpha} &=& 6\,Y_{\Delta Q}+ \sum_i\left(Y_{\Delta
    u_i}+Y_{\Delta d_i}\right)- 3\,(2Y_{\Delta \ell_\alpha}+Y_{\Delta
  e_\alpha})-2\,Y_{\Delta\tilde g}\,,   \\
Y_{\Delta R_B}&=&
- 6 Y_{\Delta Q} - 
\sum_i \left(13\,Y_{\Delta u_i}-5\,Y_{\Delta d_i}\right)
\nonumber \\
&& \hspace{1.2cm}
+ \sum_\alpha\left(10\,Y_{\Delta \ell_\alpha}+7\,Y_{\Delta e_\alpha}\right)
+68\,Y_{\Delta\tilde g} 
+10\,Y_{\Delta \tilde H_d}-2\,Y_{\Delta \tilde H_u},  
\label{eq:YDeltaRB}
\\
Y_{\Delta R_\chi}&=&  3\,\left(3\,Y_{\Delta u}-2\,Y_{\Delta \tilde
  g}\right)+\frac{1}{3}\, Y_{\Delta R_3},  
\label{eq:YDeltaRchi}
\end{eqnarray}
where,  in the last expression, 
\bea
\nonumber 
Y_{\Delta R_3}&=&
- 18 Y_{\Delta Q} - 
3\, \sum_i \left(11\,Y_{\Delta u_i}-4\,Y_{\Delta d_i}\right)
\nonumber \\
&& \hspace{1.2cm}
+ \sum_\alpha\left(16\,Y_{\Delta \ell_\alpha}+13\,Y_{\Delta e_\alpha}\right)
+82\,Y_{\Delta\tilde g} 
+16\,Y_{\Delta \tilde H_d}-14\,Y_{\Delta \tilde H_u}.   
\label{eq:YDeltaR3}
\eea

\noindent 
The density asymmetries of the five charges in
\Eqns{eq:YDeltaAlpha}{eq:YDeltaRchi} then define the basis
$Y_{\Delta_a}=\left\{Y_{\Delta_\alpha},Y_{\Delta R_B},Y_{\Delta
    R_\chi} \right\}$ 
in terms of which the five fermionic
density-asymmetries 
$Y_{\Delta\psi_a}=
\{Y_{\Delta\ell_\alpha},\,Y_{\Delta\tilde g},\,Y_{\Delta \tilde
    H_u}\}$, 
that are the relevant ones for the soft leptogenesis
processes,  have to be expressed. We will do this by introducing a
$5\times 5$ $A$-matrix defined according to:
\be
\label{eq:A5x5}
Y_{\Delta\psi_a} = A_{ab}\,  Y_{\Delta_b}\,, 
\ee
where the numerical values of $A_{ab}$ are obtained from
\Eqns{eq:YDeltaAlpha}{eq:YDeltaRchi} subjected to the constraining
conditions listed in items 3 to 5.  Let us note at this point
that the $3\times 5$ submatrix $A_{\ell_\alpha b}$ for the lepton densities
represents the generalization of the $A$ matrix introduced
in~\cite{barbieri}, $A_{\tilde H_u b}$ generalizes the Higgs
$C$-vector first introduced in~\cite{spectator2}, and $A_{\tilde g b}$
generalizes the $C$-vector for the gauginos first introduced
in~\cite{ournse}. As regards the density asymmetries for the bosonic
partners of $\ell_\alpha$ and of $\widetilde H_u$, they are simply given
by: $A_{\widetilde{\ell}_\alpha\, b}=2 \left(A_{\ell_\alpha\, b} +
  A_{\widetilde{g} b}\right)$ and $A_{H_{u}\, b} = 
2\left(A_{\widetilde{H}_u}+A_{\widetilde{g}\, b}\right)$.

\medskip

\subsection{Case I: Electron and down-quark  Yukawa reactions in equilibrium} 
\label{sec:edin}

If the down-type Higgs vev is relatively small $v_d\ll v$, the values
of the electron and down-quark masses are obtained for correspondingly
large values of the $h_d$ and $h_e$ Yukawa couplings.  For
$v_u/v_d=\tan\beta \gsim 35$ we have a regime in which at $T \sim
10^8\,$GeV, that is well above the NSE threshold Eq.~\eqref{eq:Tgmu},
both $h_d$ and $h_e$ related reactions are in equilibrium.  In this
case all the eight Yukawa conditions~\Eqns{eq:leptons}{eq:upquarks}
hold. Solving for the densities-asymmetries $Y_{\Delta\psi_a}=
\{Y_{\Delta\ell_\alpha},\,Y_{\Delta\tilde g},\,Y_{\Delta \tilde
  H_u}\}$ in terms of the charge-asymmetries
$Y_{\Delta_a}=\left\{Y_{\Delta_\alpha},Y_{\Delta R_B},Y_{\Delta
    R_\chi} \right\}$ subject to the constraints in items  3 to
5, yields 
\begin{equation}
A=\frac{1}{9\times 827466}\left(
\begin{array}{rrrrr}
 -788776 &  38690 &  38690 & -56295 &  41931  \\
   38690 &-788776 &  38690 & -56295 &  41931  \\
   38690 &  38690 &-788776 & -56295 &  41931  \\
   41913 &  41913 &  41913 & 124281 &  12798  \\
 -102411 &-102411 &-102411 & 108108 &-335907
\end{array}
\right).  
\label{eq:Aedin}
\end{equation}
%

\medskip

\subsection{Case II: Electron and down-quark Yukawa reactions out of 
equilibrium}
\label{sec:edout}
If $v_d$ is not much smaller than $v_u$, resulting in $\tan\beta \lsim
5$, then both $h_e$ and $h_d$ are sufficiently small that at $T \sim
10^8\,$GeV the related Yukawa reactions do not occur.  In this case we
have to set $h_d,\,h_e\to 0$ and the corresponding two Yukawa
equilibrium conditions in~\Eqns{eq:leptons}{eq:downquarks} do not
hold. However, two conservation laws replace these conditions.
$h_e\to 0$ implies that we gain a `chiral' symmetry for the
right-handed fermion and scalar electrons, ensuring that the total
number-density asymmetry $\Delta n_{e}+\Delta n_{\tilde e}$ is
conserved. As usual, we assume that the constant value of this
quantity vanishes, which in terms of the rescaled density asymmetries
per degree of freedom implies:
\begin{equation}
  \label{eq:YeR}
  Y_{\Delta e}-\frac{2}{3}\, Y_{\Delta \tilde g}=0\,.
\end{equation}
For the right-handed down quark we could define an anomaly-free charge 
completely equivalent to $Y_{\Delta R_\chi}$ in~\Eqn{eq:YDeltaRchi}  
but, given that in this regime all the dynamical equations  
are symmetric  under the exchange $u \leftrightarrow d$, it is
equivalent, and much more simple, to impose the condition
\begin{equation}
  \label{eq:YdR}
  Y_{\Delta d}=  Y_{\Delta u}\,.
\end{equation}
The net result is that, with respect to the previous case, the total
number of constraints is not changed, and again five quantities
suffice to express the rescaled density asymmetries for all the
fields. For the $5\times 5$ $A$ matrix defined in~\Eqn{eq:A5x5}
we obtain:
\begin{equation}
A=\frac{1}{9\times 162332}\left(
\begin{array}{rrrrr}
-210531 &  21573 &  21573 & -12414 & 12483 \\
   8676 &-165529 &  -3197 & -17958 & 29709 \\
   8678 &  -3197 &-165529 & -17958 & 29709 \\
   7497 &   7299 &   7299 &  23634 &  4833 \\
 -11322 & -18477 & -18477 &  23940 &-74385
\end{array}
\right). 
\label{eq:Aedout}
\end{equation}
%


\section{Basic Boltzmann Equations}
\label{sec:BE}
In order to render clear the role played by the new charges $\Delta
R_B$ and $\Delta R_\chi$ and by NSE effects, in this section we
introduce a simplified set of BE including only decays and inverse
decays of heavy neutrinos and sneutrinos.  
However, for the numerical results that are 
discussed in the next section, we have used the  
more complete (and involved) set of equations
described in Appendix~\ref{Appendix-B}. 

The evolution of the number density of the heavy states normalized to
the entropy density $s$ is given by
\begin{eqnarray}
\dot Y_{N} & \!= \! &
-\left(\frac{Y_{N}}{Y_{N}^{eq}}-1\right)
\gamma_{N}\,,
\label{eq:DBE_N}
\qquad\qquad
\mbox{\hspace{1.2cm} \phantom{i}} \\
\label{eq:DBE_tildeN}
\dot Y_{\widetilde{N}}
& \! =\! & 
-\left(\frac{Y_{\widetilde{N}}}{Y_{\widetilde{N}_+}^{eq}}-2\right) 
\frac{1}{2}\gamma_{\widetilde{N}}\,, 
\end{eqnarray}
where the time derivative is defined as $\dot Y= sHz\frac{dY}{dz}$
with $z=M/T$, and $H=H(z)$ is the Hubble parameter. In~\Eqn{eq:DBE_N}
$\gamma_{N}$ represents the (thermal averaged) total decay width of
the heavy neutrino $N$ into particles and sparticles of all
$\alpha$-flavours $\gamma_{N}=\sum_\alpha \gamma_{N}^\alpha$ and
$Y_{N}^{eq}$ the $N$ equilibrium density.  For the heavy sneutrinos,
we denote with $\widetilde{N}$ the sum of $\widetilde{N}_+$ and
$\widetilde{N}_-$. Thus in \Eqn{eq:DBE_tildeN} $ Y_{\widetilde{N}}=
Y_{\widetilde{N}_+}+ Y_{\widetilde{N}_-}$, while
$Y_{\widetilde{N}_+}^{eq}$ represents the equilibrium density of a
single sneutrino. For the reaction rates we have
\begin{equation}
  \label{eq:gammatN}
  \gamma_{\widetilde{N}} = \gamma_{\widetilde{N}_+}+\gamma_{\widetilde{N}_-}=
\sum_{p=s,f}\sum_\alpha\left(
\gamma_{\widetilde{N}_+}^{p\,\alpha}+
\gamma_{\widetilde{N}_-}^{p\,\alpha}\right)\,,
\end{equation}
where  the $p$ sum in the r.h.s 
of the last equality is 
over $s$-scalars and $f$-fermions final states,  
while $\gamma_{\widetilde{N}_+}=\gamma_{\widetilde{N}_-}
=\gamma_{\widetilde{N}}/2$.

In writing down the evolution equations for the five charges
$Y_{\Delta_\alpha},Y_{\Delta R_B},Y_{\Delta R_\chi}$ it is convenient
to introduce a special notation for the scalars and fermions density
asymmetries (per degree of freedom) normalized to the respective
equilibrium densities $Y^{eq}_s = 2 Y^{eq}_f= \frac{15}{4\pi^2 g^*}$:
\begin{equation}
  \label{eq:calY}
{\cal Y}_{\Delta s,\Delta f} \equiv \frac{Y_{\Delta s,\Delta f}}{Y^{eq}_{s,f}}.  
\end{equation}
Using~Eqs. \eqref{eq:Dnmu} and   \eqref{eq:Y} together   
with \eqref{eq:tQtell} and \eqref{eq:HuHd} it is then easy to 
verify that
\begin{equation}
  \label{eq:Ytg}
  {\cal Y}_{\Delta \tilde \ell,\Delta H_u}=  
{\cal Y}_{\Delta \ell,\Delta \tilde H_u}
+ {\cal Y}_{\Delta \tilde g}\,. 
\end{equation}
Including only decays and inverse decays, the Boltzmann equation for
the flavour charges $\Delta_\alpha=B/3-L_\alpha$ read:
\begin{eqnarray}
\dot Y_{\Delta\alpha}&=& 
- \epsilon_f^\alpha\left(z\right)
\left(\frac{Y_{\widetilde{N}}}{Y_{\widetilde{N}_+}^{eq}}-2\right)
\frac{\gamma_{\widetilde{N}}}{2}
+\left({\cal Y}_{\Delta \ell_\alpha}
+{\cal Y}_{\Delta \widetilde H_u}\right)
\frac{\gamma_{\widetilde{N}}^{f,\alpha}}{2}
+\Big({\cal Y}_{\Delta \ell_\alpha}+{\cal Y}_{\Delta H_u}\Big)
\frac{\gamma_{N}^\alpha}{4}  
\nonumber \\
&&-\epsilon_s^\alpha\left(z\right)
\left(\frac{Y_{\widetilde{N}}}{Y_{\widetilde{N}_+}^{eq}}-2\right)
\frac{\gamma_{\widetilde{N}}}{2}
+\left({\cal Y}_{\Delta \widetilde \ell_\alpha}
+{\cal Y}_{\Delta H_u}\right)\frac{\gamma_{\widetilde{N}}^{s,\alpha}}{2}
+\left({\cal Y}_{\Delta \widetilde \ell_\alpha}
+{\cal Y}_{\Delta \widetilde H_u}\right)\frac{\gamma_{N}^\alpha}{4}\,.
\end{eqnarray}
To write this expression in a more compact form, we define the total
flavoured CP asymmetry $\epsilon^\alpha= \epsilon^\alpha_f+
\epsilon^\alpha_s$ and the total sneutrinos decay rate into $\alpha$
leptons and sleptons
$\gamma_{\widetilde{N}}^{\alpha}=\gamma_{\widetilde{N}}^{f,\alpha}+
\gamma_{\widetilde{N}}^{s,\alpha}$. 
For quantities without a flavour index a sum over flavour 
will be understood, e.g.:
$\gamma_{\widetilde{N}}=\sum_\alpha
\gamma_{\widetilde{N}}^{\alpha}$ and  
$\epsilon_{f,s}=\sum_\alpha\epsilon_{f,s}^\alpha$. 
Furthermore, we can
use~\Eqn{eq:Ytg} to express the density asymmetries of the scalars in
terms of the ones of the fermions, and to an excellent approximation
we can write $\gamma_{\widetilde{N}}^{s,\alpha}=
\gamma_{\widetilde{N}}^{f,\alpha}$.\footnote{ For $M\sim 10^8\,$GeV,
  the soft terms corrections to this approximation
  $\gamma_{\widetilde{N}}^{s}/
  \gamma_{\widetilde{N}}^{f}-1=(A^2-AB)/M^2$ can be safely
  neglected.}  After the same notational simplifications are applied
also to the BE for $Y_{\Delta R_B}$ and $Y_{\Delta
  R_\chi}$, the following set is obtained:
\begin{eqnarray}
  \dot Y_{\Delta\alpha}&=&
  - \epsilon^\alpha\left(z\right)
  \left(\frac{Y_{\widetilde{N}}}{Y_{\widetilde{N}_+}^{eq}}-2\right)
  \frac{\gamma_{\widetilde{N}}}{2}+
  \left({\cal Y}_{\Delta \ell_\alpha}
    +{\cal Y}_{\Delta \widetilde H_u}+{\cal Y}_{\Delta \widetilde g}
  \right)
  \frac{\gamma_{N}^\alpha+\gamma_{\widetilde{N}}^\alpha}{2}\,,   
  \label{eq:DeltaAlpha}
\\
\dot Y_{\Delta R_B}&=&
 \epsilon_s\left(z\right)
\left(\frac{Y_{\widetilde{N}}}{Y_{\widetilde{N}_+}^{eq}}-2\right)
\gamma_{\widetilde{N}}-
\sum_\alpha 
\left(
{\cal Y}_{\Delta \ell_\alpha}+
{\cal Y}_{\Delta \widetilde H_u}+{\cal Y}_{\Delta \widetilde g}
\right)
\frac{\gamma_{N}^\alpha+\gamma_{\widetilde{N}}^\alpha}{2}
- 
{\cal Y}_{\Delta \widetilde g}
\frac{\gamma_{\widetilde{N}}}{2}
\,, \ \ 
  \label{eq:DeltaRB}
\\
\dot Y_{\Delta R_\chi}&=& 
\left[\epsilon_s\left(z\right)- 
\epsilon_f\left(z\right)\right]
\left(\frac{Y_{\widetilde{N}}}{Y_{\widetilde{N}_+}^{eq}}-2\right)
\frac{\gamma_{\widetilde{N}}}{6}
- {\cal Y}_{\Delta \widetilde g}\,
\frac{\gamma_{\widetilde{N}}}{6}\,.  
  \label{eq:DeltaRchi}
\end{eqnarray}
It is possible, and formally straightforward, to add to these
equations the appropriate terms that allow to extend their validity
also in the SE regime, that is for sneutrino masses below the
bound~\Eqn{eq:NSE}. In order to do this, we denote by $\gamma_{\tilde
  g}^{\rm eff}$ the set of gaugino-mediated reactions with chirality
flip on the gaugino line that are responsible for processes that
equilibrate particle-sparticle chemical
potentials.\footnote{Ref.~\cite{Plumacher:1997ru} included a similar
  term $\gamma_{\rm MSSM}$ in the BE for supersymmetric leptogenesis,
  corresponding to the thermally averaged cross section for the
  photino mediated process $e+e\leftrightarrow
  \widetilde{e}+\widetilde{e}$ computed
  in~\cite{Keung:1983nq}. However, in the total cross section the only
  contributions that do not vanish in the $m_{\tilde \gamma}\to 0$
  limit are those that, like e.g.  $e^-_L+e^-_R \leftrightarrow
  \widetilde{e}_L+\widetilde{e}_R$, do not enforce
  SE. Superequilibrating reactions like $e^-_L+e^-_L \leftrightarrow
  \widetilde{e}_L+\widetilde{e}_L$ all vanish in the $m_{\tilde
    \gamma}\to 0$ limit.} We also denote by $\gamma^{\rm
  eff}_{\submuh}$ the set of reactions induced by the higgsino mixing
parameter $\muh$ that enforce the chemical equilibrium
condition $\tilde H_u+\tilde H_d=0$. The thermally averaged rates for
these reactions can be written in an approximated form as:
\begin{equation}
  \label{eq:approxSE}
\frac{\gamma_{\tilde g}^{\rm eff}}{n^{eq}_f} = \frac{m^2_{\widetilde g}}{T}, 
\qquad\qquad 
\frac{\gamma_{\submuh}^{\rm eff}}{n^{eq}_f} = \frac{\muh^2}{T}, 
\end{equation}
where $n_f^{eq}$ is the equilibrium number density for one fermionic
degree of freedom, while $m_{\widetilde g }$ and $\muh$ in these
equations have to be understood as effective mass parameters in which
all coupling constants as well as reaction multiplicities are
reabsorbed.  Extension of the validity of
\Eqns{eq:DeltaAlpha}{eq:DeltaRchi} to the SE domain is then achieved
by adding the following terms to the equations for $R_B$ and $R_\chi$:
\begin{eqnarray}
  \label{eq:DeltaRBSE}
\dot Y_{\Delta R_B}^{SE}&=&\left\{\dot Y_{\Delta R_B}\right\}  
- {\cal Y}_{\Delta \widetilde g}\>\gamma_{\tilde g}^{\rm eff}\,,
\\ \label{eq:DeltaRchiSE}
\dot Y_{\Delta R_\chi}^{SE}&=&
\left\{\dot Y_{\Delta R_\chi}\right\}  
- \frac{1}{3}{\cal Y}_{\Delta \widetilde g}\> \gamma_{\tilde g}^{\rm eff}
+ \frac{1}{3}\left({\cal Y}_{\Delta \widetilde H_u}+ 
{\cal Y}_{\Delta \widetilde H_d}\right)\,\gamma_{\submuh}^{\rm eff}\,,
\end{eqnarray}
where the $\big\{\dot Y_{\Delta R}\big\}$ above stand for the r.h.s of
the corresponding equations \eqref{eq:DeltaRB} and
\eqref{eq:DeltaRchi}. Note that since the $R_B$ charge of the $\muh$
term is $R_B(H_uH_d)=2$, $\muh$ conserves $R_B$ and accordingly there
is no term proportional to $\gamma_{\submuh}^{\rm eff}$ in
\Eqn{eq:DeltaRBSE}.  Since higgsino equilibration involves also the
density asymmetry ${\cal Y}_{\Delta \widetilde H_d}$ we give below the
corresponding $C$ vectors to express it in terms of the basis of the
charge-asymmetries:
\begin{eqnarray}
  \label{eq:C_I}
{\rm Case\ I:}\qquad\  C^{\tilde H_d}&=&  
\frac{1}{827466}\left(14237,\;
14237,\;14237,\;1260,\;-3915\right)\,,
\\  \label{eq:C_II}
{\rm Case\ II:}\qquad C^{\tilde H_d}&=&  
\frac{1}{3\times 162332  }\left(12469,\;
16768,\;16768,\;7056,\;-21924\right)\,.
\end{eqnarray}
We have of course checked that by increasing the values
of $m_{\widetilde g }$ and $\muh$, the results of integrating the 
set of BE given by 
\Eqn{eq:DeltaAlpha} and \Eqns{eq:DeltaRBSE}{eq:DeltaRchiSE}
converge to the solutions of the usual BE for the SE regime
(see Appendix~\ref{Appendix-B}).

\medskip

\subsection{NSE Regime: R-genesis in a simple case}
\label{sec:simple}

To highlight the role played by the asymmetries of the two $R$
charges, let us define a simple scenario, in which lepton flavour
effects play basically no role and thus do not shadow the new effects.
This scenario is defined by the following two conditions:

\begin{itemize}
\item We assume equal branching fractions for the decays of $N$ and of
  $\widetilde{N}_\pm$ into the three lepton flavours, that is the
  $P_\alpha$ defined in \Eqn{eq:Palpha} are all equal to
  $\frac{1}{3}$ implying $\epsilon^\alpha= \frac{1}{3}\epsilon$ and $
  \gamma^\alpha_{N,\widetilde{N}} =\frac{1}{3} \gamma_{N,\widetilde{N}}$.

\item We assume the regime described in Case I,
  Section~\ref{sec:edin}, in which the 
Yukawa equilibrium condition for the electron holds, 
and thus the three lepton flavours are 
all treated on equal footing (see the $3\times 3$ 
upper-left corner in the $A$-matrix 
\Eqn{eq:Aedin}). Given the previous 
condition, it is then useful to define  a `flavour averaged' 
lepton density asymmetry as: 

\begin{equation}
  \label{eq:Yell}
{\cal Y}_{\Delta \ell} =\frac{1}{3} \sum_\alpha
{\cal Y}_{\Delta \ell_\alpha}  
\end{equation}

\end{itemize}

With these conditions, the three equations for the flavour
charges~\Eqn{eq:DeltaAlpha} can be resummed in closed form into a
single equation for $B-L$:
\begin{equation}
\label{eq:DeltaB-LS}
\dot Y_{\Delta_{B-L}}=
  - \epsilon\left(z\right)
  \left(\frac{Y_{\widetilde{N}}}{Y_{\widetilde{N}_+}^{eq}}-2\right)
  \frac{\gamma_{\widetilde{N}}}{2}+
\left({\cal Y}_{\Delta \ell}
    +{\cal Y}_{\Delta \widetilde H_u}+{\cal Y}_{\Delta \widetilde g}
  \right)
  \frac{\gamma_{N}+\gamma_{\widetilde{N}}}{2}\,,    
\end{equation}
yielding a reduced set of just 3 BE.
The $3\times 3$ matrix relating
$\{Y_{\Delta\ell},\,Y_{\Delta\tilde g},\,Y_{\Delta \tilde H_u}\}$ to
the three charge-asymmetries $\left\{Y_{\Delta_{B-L}},Y_{\Delta
    R_B},Y_{\Delta R_\chi} \right\}$
can be readily evaluated
from \Eqn{eq:Aedin}:
\begin{equation}
A=\frac{1}{827466}\left(
\begin{array}{rrrrr}
  -26348 & -6255 &  4659  \\
    4657 & 13809 &  1422  \\
  -11379 & 12012 &-37323
\end{array}
\right).  
\label{eq:Aedin3x3}
\end{equation}
It is now easy to see that in the NSE regime we can rewrite the BE  as
\begin{eqnarray}
  \dot Y_{\Delta_{B-L}}
&=&3\,\dot Y_{\Delta R_\chi}-\dot Y_{\Delta R_B}\,,   
  \label{eq:DeltaB-LnoS}
\\
\dot Y_{\Delta R_B}
&=& \epsilon_s(z)\,
\left(\frac{Y_{\widetilde{N}}}{Y_{\widetilde{N}_+}^{eq}}-2\right)
{\gamma_{\widetilde{N}}}
-
\left(
{\cal Y}_{\Delta \ell}+
{\cal Y}_{\Delta \widetilde H_u}+{\cal Y}_{\Delta \widetilde g}
\right)
\frac{\gamma_{N}+\gamma_{\widetilde{N}}}{2}
- 
{\cal Y}_{\Delta \widetilde g}
\frac{\gamma_{\widetilde{N}}}{2} \,, 
\label{eq:DeltaRBS} 
\\
\dot Y_{\Delta R_\chi}&=& 
\left[\epsilon_s\left(z\right)- 
\epsilon_f\left(z\right)\right]
\left(\frac{Y_{\widetilde{N}}}{Y_{\widetilde{N}_+}^{eq}}-2\right)
\frac{\gamma_{\widetilde{N}}}{6}
- {\cal Y}_{\Delta \widetilde g}
\frac{\gamma_{\widetilde{N}}}{6} \,, 
\label{eq:DeltaRchiS}
\end{eqnarray}
since the difference in the r.h.s. of \Eqn{eq:DeltaB-LnoS} gives
precisely \Eqn{eq:DeltaB-LS}.  \Eqn{eq:DeltaB-LnoS} makes apparent how
$Y_{\Delta R_\chi}$ and $Y_{\Delta R_B}$, that in the $T\to 0$ limit
keep having non vanishing CP asymmetries, are sources of the $B-L$
asymmetry.  This result is in fact completely general: the only role
of the two conditions listed above is simply that of allowing to
collapse the three equations for $\Delta_\alpha$ into a single one for
$\Delta_{B-L}$, while maintaining the BE equations in closed form. In
particular, it holds also when scattering processes are included (see
Appendix~\ref{Appendix-B}) and independently of the particular (NSE)
temperature regime and flavour configuration. In short, in the NSE
regime the evolution of $\Delta_{B-L}$ can be always obtained from the
evolution of $3\Delta_{R_\chi}- \Delta_{R_B}$, and the final value of
$Y_{\Delta_{B-L}}$ can be equally well obtained from summing the
values of the flavour charges asymmetries $\sum_\alpha
Y_{\Delta_\alpha}$ or from the final value of $3Y_{\Delta R_\chi}-
Y_{\Delta R_B}$.  The reason why this happens is simple: by using the
definitions~\Eqns{eq:RB}{eq:chiralup} together with
\Eqns{eq:R2}{eq:R3} one obtains that $3R_\chi- R_B= \raise 2pt
\hbox{$\chi$}_{u^c_L} -\frac{2}{3}B-PQ $. Of course, only the $PQ$
fragment of this charge is violated in sneutrinos interactions, and
from Table~\ref{tab:1} we see that this violation is precisely the
same than for $B-L$ (e.g. for $\widetilde N \to \ell \tilde H_u$ we
have $\Delta(B-L)=-\Delta L=-\Delta (PQ)=-1$). Thus, regardless of the
fact that $B-L$, $R_B$ and $R_\chi$ are all independent charges, in
the NSE regime the BE for $3Y_{\Delta R_\chi}- Y_{\Delta R_B}$ will
always coincide with the BE for $Y_{\Delta_{B-L}} =\sum_\alpha
Y_{\Delta_\alpha}$.

In our particularly simple case we can take a further step.  Let us
rewrite the density asymmetry ${\cal Y}_{\Delta \widetilde g}$ and the
combination $({\cal Y}_{\Delta \ell}+ {\cal Y}_{\Delta \widetilde
  H_u}+{\cal Y}_{\Delta \widetilde g})$ in the r.h.s of
\Eqns{eq:DeltaRBS}{eq:DeltaRchiS} in terms of
$Y_{\Delta_{B-L}},\,Y_{\Delta R_B},\,Y_{\Delta R_\chi}$ by means of
the $A$ matrix \Eqn{eq:Aedin3x3}. We can then 
replace
$Y_{\Delta_{B-L}}\to 3 Y_{\Delta R_\chi} - Y_{\Delta R_B}$ and, 
by using $\gamma_N =\gamma_{\widetilde N}$ 
we obtain:
\begin{eqnarray}
\label{eq:DeltaRchiS2}
3 \dot Y_{\Delta R_\chi}&=& 
\left[\epsilon_s\left(z\right)- 
\epsilon_f\left(z\right)\right]
\left(\frac{Y_{\widetilde{N}}}{Y_{\widetilde{N}_+}^{eq}}-2\right)
\frac{\gamma_{\widetilde{N}}}{2}-
\frac{1}{827466}
\left(9152\,Y_{\Delta R_B} 
+15393\, Y_{\Delta R_\chi}\right)
\frac{\gamma_{\widetilde{N}}}{2}
\,,\qquad 
\\
\dot Y_{\Delta R_B}
&=& 2\,\epsilon_s(z)\,
\left(\frac{Y_{\widetilde{N}}}{Y_{\widetilde{N}_+}^{eq}}-2\right)
\frac{\gamma_{\widetilde{N}}}{2}\
-\ \frac{1}{827466}
\left(114424\,Y_{\Delta R_B} 
-245511\, Y_{\Delta R_\chi}\right)
\frac{\gamma_{\widetilde{N}}}{2}
\,. \ \ 
\label{eq:DeltaRBS2}
\end{eqnarray}
These two equations show that although the asymmetries produced in the
two charges $3 R_\chi$ and $R_B$ tend to cancel when taking the
difference, their respective washouts are quite different, and such a
cancellation will never occur.  In the general case flavour dynamics
does not allow to collapse the set of BE to just two equations, but
still the same mechanism is at work: because of the different
washouts, the difference between $3 Y_{R_\chi}$ and $Y_{R_B}$ becomes
of the same order of these density asymmetries, and so does
$Y_{\Delta_{B-L}}$.  Perhaps surprisingly, we can then expect that by
increasing the washouts from a strength of order weak up to (not too)
large strengths, the final value of $B-L$ will increase.  The
numerical results in the next section confirm this picture.

In the SE regime instead, things proceed in a different way.
\Eqns{eq:DeltaRBSE}{eq:DeltaRchiSE} show that the BE for $Y_{R_\chi}$
and $Y_{R_B}$ acquire new washout terms, that are proportional to the
SE rates, while on the contrary no analogous terms enter the BE for
$Y_{\Delta_\alpha}$ \Eqn{eq:DeltaAlpha} or for $Y_{\Delta_{B-L}}$
\Eqn{eq:DeltaB-LS}.  Thus, in the SE regime, \Eqn{eq:DeltaB-LnoS} does
not hold. One can argue instead that, because of the SE washouts, the
roles of $\Delta_{B-L}$ and of $3\Delta R_\chi-\Delta R_B$ get
reversed, since now we have
\begin{equation}
3\,\dot Y_{\Delta R_\chi}-\dot Y_{\Delta R_B}=
\dot Y_{\Delta_{B-L}}+
\left({\cal Y}_{\Delta \widetilde H_u}+ 
{\cal Y}_{\Delta \widetilde H_d}\right)
\,\gamma_{\submuh}^{\rm eff}\,.
\end{equation}
In other words, since SE reactions conserve $B-L$ but violate the $R$
and $PQ$ charges, the only source of asymmetry surviving SE is the
(thermally induced) $Y_{\Delta_{B-L}}$ asymmetry. Given that
$\Delta{R_\chi}$ and $\Delta{R_B}$ both contain `fragments' that carry
$B$ number, they do not vanish in the SE limit, but are driven to
values that are proportional to $\Delta_{B-L}$.  The constants of
proportionality are determined by the chemical equilibrium and
conservation law conditions appropriate for the specific regime and,
for example, in Case I are given by $Y_{\Delta
  {R_B}}=-\frac{1}{3}Y_{\Delta_{B-L}}$ and $Y_{\Delta
  {R_\chi}}=-\frac{3}{79} \,Y_{\Delta_{B-L}}$.

\medskip

\section{Results}
\label{sec:results}

In this section we quantify the results that are obtained for the
baryon asymmetry yield of soft leptogenesis when the effective theory
described in the previous sections is used, and we confront them with
what is obtained in the standard scenario, in which SE is assumed to
hold at all temperatures.  Our results are obtained by numerical
integration of the BE given in Appendix~\ref{Appendix-B} that also
include various scattering processes.  The comparative results for the
SE case can be obtained in two formally different, but physically
equivalent, ways.  A first possibility is that of taking the limit
$m_{\widetilde g},\muh \to \infty$ in the complete BE (given, for
example, in their basic form in~\Eqns{eq:DeltaRBSE}{eq:DeltaRchiSE}).
A second possibility, that corresponds to usual treatments, is to
solve only the three equations for the flavour charge-density
asymmetries $Y_{\Delta_\alpha}$ with the corresponding $A$ matrix and
$C$ vectors obtained under the assumption of SE.  For the two cases we
are analyzing: Case I ($h_{e,d}$ Yukawa equilibrium) and Case II
($h_{e,d}$ Yukawa non-equilibrium), we give the corresponding matrices
in Appendix~\ref{Appendix-B} in \Eqns{eq:ACTin}{eq:ACTout}. Of course,
we have verified that both procedures yield the same results.

Some of our results are presented in terms of an efficiency
parameter $\eta$ defined according to:
\begin{equation}
\label{eq:eta}
\eta\equiv \left|\frac{Y_{\Delta_{B-L}}(z\rightarrow \infty)}
{2\, \bar\epsilon \, Y^{\rm eq}_{\widetilde N}(z\to 0)}\right| 
\end{equation}
where $\bar\epsilon$ is defined in Eq.\eqref{eq:eptot}.  To single out
the new effects that we want to quantify, all our results are obtained
assuming a configuration of flavour equipartition, with all the
flavour branching fractions~\Eqn{eq:Palpha} equal: $P_\alpha =
\frac{1}{3}$, so that flavour effects are basically switched off.  In
all cases, the heavy sneutrino mass is held fixed at $M=10^8\,$GeV,
that is above the temperature threshold for the validity of the
effective theory~\Eqn{eq:Tgmu}. The values of the other relevant
parameters are: $A=1\,$TeV, $\phi_A=\frac{\pi}{2}$ and
$\bar\epsilon=\frac{A}{M}=10^{-5}$ that corresponds to a resonantly
enhanced CP asymmetry in mixing.  This is obtained for $2\,B\sim
\Gamma\sim 2.6\,\left(\frac{m_{\rm eff}}{0.1\,{\rm eV}}\right)\,$GeV.
As regards gaugino mass dependent contributions to the CP asymmetries
from vertex corrections, as was mentioned in
Section~\ref{sec:motivations} they are suppressed by additional powers
of $\Lambda_{susy}/M$.  Given the large value of $M$ that we are
using, they remain irrelevant even in the cases labeled as the
``$m_{\tilde g}\to \infty$ limit'', since in practice $m_{\tilde
  g}\approx 10\,$TeV is more than sufficient to enforce SE, and this
is the value we are effectively using. Therefore, in our regime $\bar
\epsilon$ is essentially determined only by $CP$ violation in mixing.

%
\begin{figure}[t!]
\includegraphics[width=\textwidth]{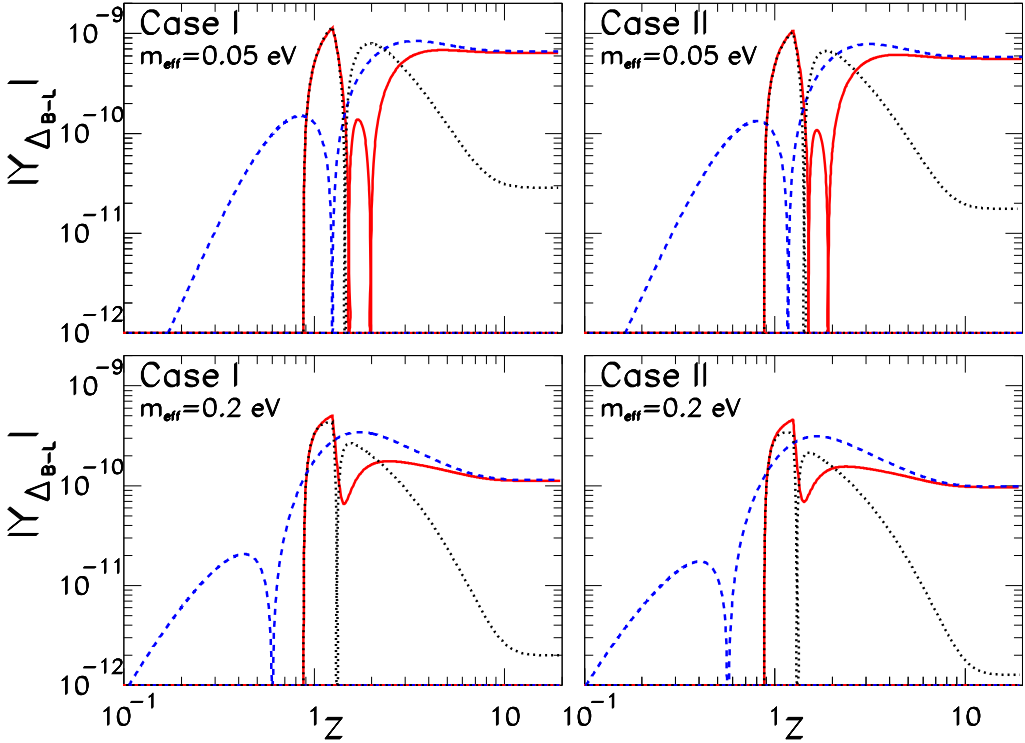}
%
%
 \caption[]{Evolution of $Y_{\Delta_{B-L}}$. The solid
   continuous (red) line depict the complete results in the $m_{\widetilde
     g}=\muh\to 0\,$ limit.  The dashed (blue) line correspond to the same
   limit but with all thermal corrections to the $CP$ asymmetries
   neglected.  The dotted (black) line gives $Y_{\Delta_{B-L}}$ with thermal
   effects when SE is assumed.  Panels on the left and right sides
   are respectively for Case I ($h_{e,d}$ Yukawa equilibrium) and Case
   II ($h_{e,d}$ Yukawa non-equilibrium).  Upper and lower panels are
   respectively for $m_{\rm eff}=0.05\,$eV and $m_{\rm eff}=0.20\,$eV.}
\label{fig:evolutions}
\end{figure} 
%

We plot in Figure~\ref{fig:evolutions} the evolution of
$Y_{\Delta_{B-L}}$ with increasing $z=M/T$.  The solid (red) lines
correspond to the full results obtained in the $m_{\widetilde
  g},\,\muh\to 0\,$GeV limit, that is when particles-sparticles
superequilibrating processes are completely switched off.  The dashed
(blue) lines give the results obtained in the same limit, but when all
thermal corrections to the CP asymmetries are neglected, and
$\epsilon_s=-\epsilon_f=\bar\epsilon/2$. From all the four panels we
see that in the NSE regime neglecting thermal corrections is an
excellent approximation that reproduces with very good accuracy the
(sizable) final values of $Y_{\Delta_{B-L}}$.  The dotted (black)
lines give $Y_{\Delta_{B-L}}$ in the usual treatments which includes
thermal corrections and also assumes SE, that in our treatment
corresponds to taking the limit $m_{\widetilde g},\muh \to \infty$.
Panels on the left side refer to Case I discussed in
Section~\ref{sec:edin}, panels on the right side are for Case II
discussed in Section~\ref{sec:edout}. We can see that the differences
between the situations in which the $h_{e,d}$ Yukawa reactions are in
equilibrium and when they are out of equilibrium are rather mild.
Therefore in the following we will concentrate just on results for
Case I.  Upper and lower panels correspond instead to two different
strength for the washout processes, parameterized respectively by
$m_{\rm eff}=0.05\,$eV and $m_{\rm eff}=0.20\,$eV.  As it was expected
from the analysis in the previous section, we see that the stronger
the washouts, the larger is the gain in efficiency with respect to the
SE scenario.

\begin{figure}[ht]
\begin{center}
\includegraphics[width=0.95\textwidth]{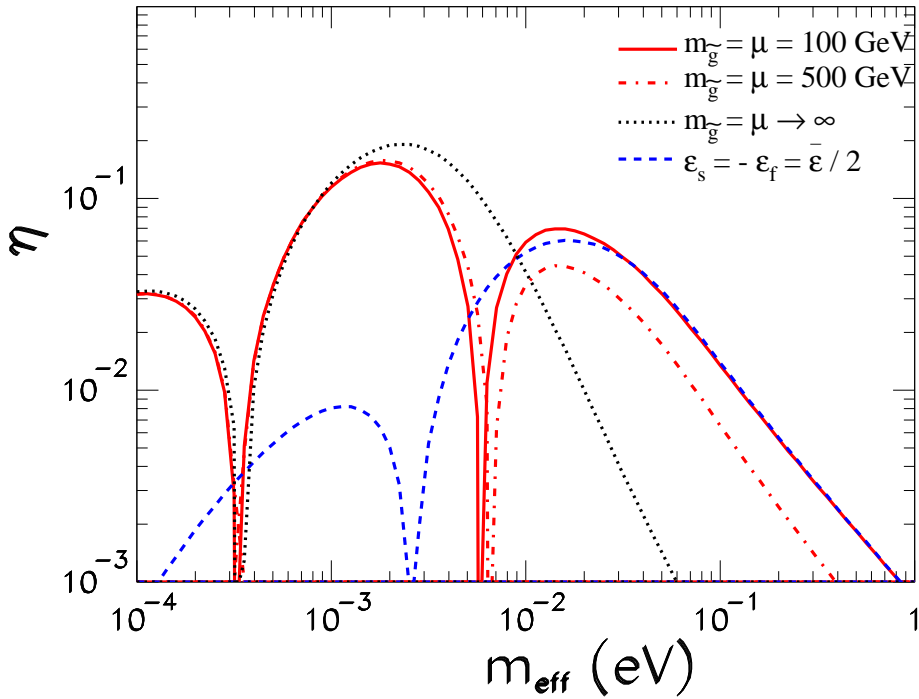}
\end{center}
\caption{Efficiency factor $\eta$ as a function of the washout
  parameter $m_{\rm eff}$ for Case I ($h_{e,d}$ Yukawa equilibrium)
  and different values of $m_{\widetilde g}=\muh$.  The red continuous
  line corresponds to the $m_{\widetilde g}=\muh=100\,$ GeV which is  
  still in the full NSE regime, while
  the dashed blue line to the same limit but with thermal
  corrections neglected.  The red dash-dotted line corresponds
  respectively to $m_{\widetilde g}=\muh= 500\,$GeV, and
  the 
  black dotted line to SE with $m_{\widetilde g},\,\muh\to \infty$.}
\label{fig:etameff}
\end{figure}

In Figure~\ref{fig:etameff} we plot for Case I the efficiency $\eta$
defined in \Eqn{eq:eta} as a function of the washout parameter $m_{\rm
  eff}$.  The red continuous line corresponds to $m_{\widetilde
  g}=\muh=100\,$GeV.  We have chosen a non-zero value for these
parameters because of phenomenological motivations, however we have
checked that the results are practically indistinguishable from those
obtained in the $m_{\widetilde g}=\muh\to 0\,$ limit and thus, in
agreement with \Eqn{eq:Tgmu}, the evolution still occurs in the full
NSE regime.  The red dash-dotted line corresponds to $m_{\widetilde
  g}=\muh=500\,$ GeV. We can see that in this case SE rates start
suppressing the efficiency, but are still far from attaining full
thermal equilibrium. The black dotted line corresponds to the
$m_{\widetilde g},\,\muh\to \infty\,$ limit of complete SE.  We see
that if SE is incorrectly assumed in temperature ranges where it does
not occur, one could vastly underestimate the leptogenesis efficiency.
The size of this underestimation is a fast increasing function of the
washouts, and for particularly large values of $m_{\rm eff}$ can reach
the two orders of magnitude level.  Let us also note that for $m_{\rm
  eff}\gsim 6\times 10^{-3}\,$eV, the assumption of SE results in a
baryon asymmetry of the wrong sign. Graphically, one can see this from
the fact that at small values of $m_{\rm eff}$ the black and red lines
approximately overlap, and then both change sign around $m_{\rm
  eff}\sim 3\times 10^{-4}\,$eV. But around $m_{\rm eff}\sim 6\times
10^{-3}\,$eV for the red line there is another sign change. This marks
the onset of $R$-genesis domination; therefore, from this point
onward, baryogenesis does not proceed through leptogenesis, but rather
through $R$-genesis.

In the same figure we have also plotted with the dash blue continuous
line the NSE results in the approximation of neglecting all thermal
corrections to the $CP$ asymmetries.  By comparing with the full
results (red continuous line) we see that for $m_{\rm eff}\gsim {\rm
  few}\,\times 10^{-2}\,$eV thermal corrections give negligible
effects. We conclude that in the case of $R$-genesis, the zero
temperature approximation yields quite reliable results.

\begin{figure}[t!]
\begin{center}
\includegraphics[width=0.95\textwidth]{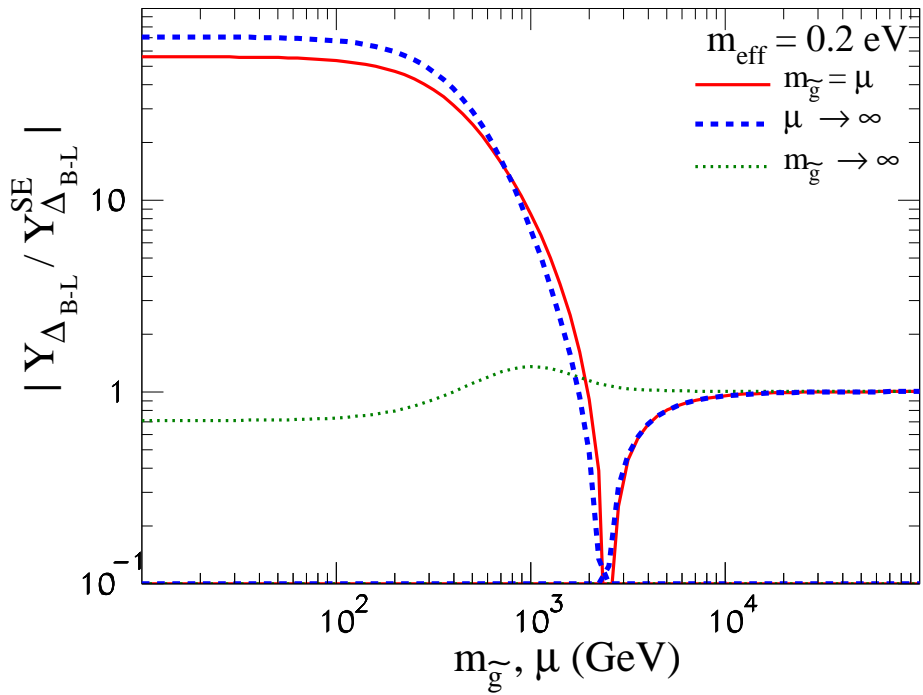}
\end{center}
\caption{The final value of $Y_{\Delta_{B-L}}$ normalized to the SE
  result $Y^{SE}_{\Delta_{B-L}}$ as a function of $m_{\widetilde g}$
  and $\muh$ for Case I ($h_{e,d}$ Yukawa equilibrium) and $m_{\rm
    eff}=0.20\,$eV.  The red continuous line corresponds to varying
  simultaneously both parameters holding $m_{\widetilde g}=\muh$.  The
  blue dashed line corresponds to varying only $m_{\widetilde g}$ in
  the limit $\muh \to \infty$.  The green dotted line corresponds to
  varying only $\muh$ in the limit $m_{\widetilde g}\to \infty$.}
\label{fig:mg-eta}
\end{figure}


In Figure~\ref{fig:mg-eta} we plot the final value of
$Y_{\Delta_{B-L}}$ as a function of different values of $m_{\widetilde
  g}$ and $\muh$, normalized for convenience to the value
$Y^{SE}_{\Delta_{B-L}}$ obtained when SE is assumed.  The results
correspond again to Case I discussed in Section~\ref{sec:edin}. In
order to enhance the impact of the new effects, we have fixed the
washout parameter to a rather large value $m_{\rm eff}=0.20\,$eV.  The
red continuous line corresponds to varying simultaneously both SE
parameters keeping their values equal: $m_{\widetilde g}=\muh$.  We
see that for $m_{\widetilde g}=\muh\lsim 1\,$TeV the amount of $B-L$
asymmetry produced by soft leptogenesis can be up to two orders of
magnitude larger (and of the opposite sign) with respect to what would
be obtained in the usual approach with SE.  SE effects start
suppressing the asymmetry around $m_{\widetilde g}=\muh\sim 1\,$TeV.
The asymmetry then changes sign around $3\,$TeV, that marks the
transition from the $R$-genesis to the leptogenesis regime, and
eventually around $5\,$TeV SE reactions attain complete thermal
equilibrium and $Y_{\Delta_{B-L}}/Y^{SE}_{\Delta_{B-L}}\to 1$.
%
\begin{figure}[t!]
\begin{center}
\includegraphics[width=0.95\textwidth]{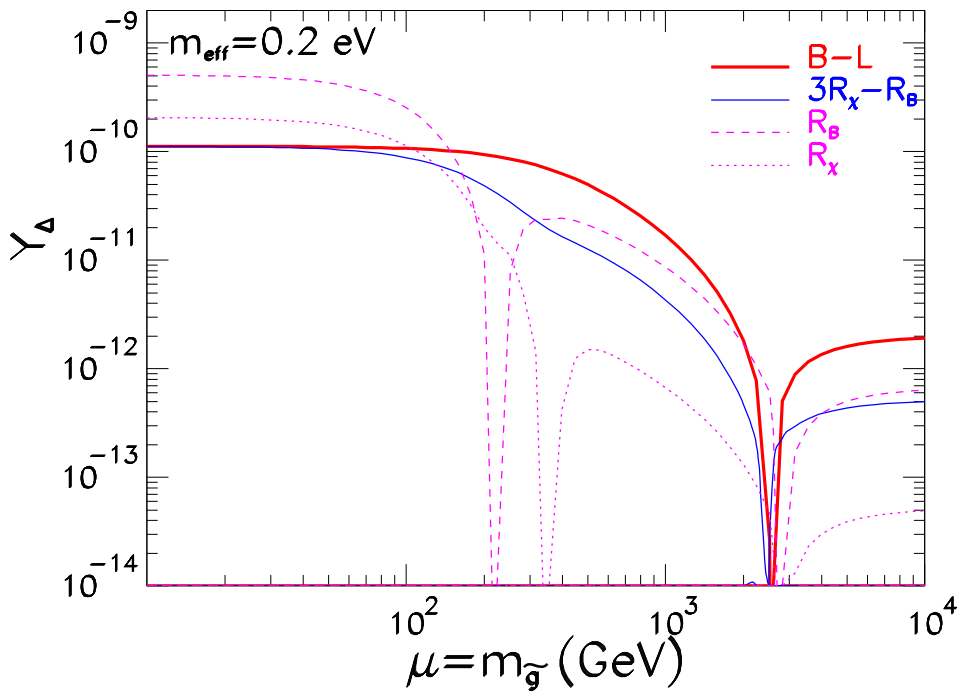}
\end{center}
\caption{Final values of the charge density asymmetries as a function
  of $m_{\widetilde g}=\muh$ for Case I ($h_{e,d}$ Yukawa equilibrium)
  and $m_{\rm eff}=0.20\,$eV.  Thick red line: $Y_{\Delta_{B-L}}$;
  thick blue line: $3Y_{\Delta R_\chi}-Y_{\Delta R_B}$; thin dashed
  purple line: $Y_{\Delta {R_B}}$; thin dotted purple line: $Y_{\Delta
    {R_\chi}}$.  }
\label{fig:rcharges}
\end{figure}
%
It can be of some interest knowing what happens 
if only one of the two anomalous symmetries $U(1)_R$ or $U(1)_{PQ}$
were present.
While we have not constructed such theories, our BE equations are
sufficiently general to allow exploring numerically also these cases.
The corresponding results are also depicted in
Figure~\ref{fig:mg-eta}.  
The blue dashed line corresponds to the $U(1)_R$-theory where 
$m_{\widetilde g}$ is varied while $U(1)_{PQ}$ is
broken.\footnote{Note that since $\muh$ breaks both symmetries, the
  case of the $U(1)_R$-theory is somewhat academic. We include it to
  put in evidence the fundamental role of $U(1)_R$ in enhancing the
  baryon asymmetry.}
The green dotted line corresponds to the alternative
$U(1)_{PQ}$-theory in which $m_{\widetilde g} \to \infty$ and only
$\muh$ is varied.
From these results we see that the real responsible of the large
effects is the $R$-symmetry, while the effects of the $PQ$ symmetry
remains qualitatively more at the level of typical spectator effects.
A theoretical justification of this behavior is not difficult to find,
and we will discuss it in the following concluding section.

Some important aspects of the transition from $R$-genesis (NSE regime)
to leptogenesis (SE regime) are highlighted in
figure~\ref{fig:rcharges}, where we plot the final value of the
relevant charge density-asymmetries as a function of $m_{\widetilde
  g}=\muh$, assuming Case~I and $m_{\rm eff}=0.20\,$eV.  The thick
solid red line corresponds to $Y_{\Delta_{B-L}}$, while the thin solid
blue line corresponds to $3Y_{\Delta R_\chi}-Y_{\Delta R_B}$.  The
thin dashed and dotted purple lines display respectively $Y_{\Delta
  {R_B}}$ and $Y_{\Delta {R_\chi}}$.  We see that up to $m_{\widetilde
  g}=\muh\sim 100\,$GeV we have $Y_{\Delta_{B-L}}\simeq 3Y_{\Delta
  R_\chi}-Y_{\Delta R_B}$ that is in agreement
with~\Eqn{eq:DeltaB-LnoS}, and thus implies that baryogenesis occurs
almost only via $R$-genesis.  As the soft supersymmetry-breaking parameters
are increased, SE reactions begin to wash out efficiently $Y_{\Delta
  {R_B}}$ and $Y_{\Delta {R_\chi}}$ but the difference $3Y_{\Delta
  {R_\chi}}-Y_{\Delta {R_B}}$ still remains of the order of
$Y_{\Delta_{B-L}}$, and $R$-genesis still gives the dominant
contribution to baryogenesis.

Around $m_{\widetilde g}=\muh\sim 3\,$TeV all the charge asymmetries
change simultaneously their sign. This is the benchmark of the onset
of the regime in which leptogenesis dominates.  The only relevant
source for generating the density-asymmetries is now the
(opposite-sign) thermally induced $B-L$ asymmetry, that is not
affected by SE washouts, and that is feeding (small) asymmetries into
all the other charges.  In this regime $Y_{\Delta_{R_B}}$ and
$Y_{\Delta_{R_\chi}}$ do not have anymore an independent dynamics, and
can be simply computed in terms of $Y_{\Delta_{B-L}}$ yielding
$Y_{\Delta {R_B}}=-\frac{1}{3}Y_{\Delta_{B-L}}$ and $Y_{\Delta
  {R_\chi}}=-\frac{3}{79} \,Y_{\Delta_{B-L}}$.


\section{Discussion and conclusions}
\label{sec:conclusions}
The supersymmetric seesaw model unavoidably entails the possibility of
soft leptogenesis. The interest in this possibility relies on the fact
that while supersymmetric leptogenesis can only proceed within
temperature regimes that are in strong tension with the bounds from
overproduction of gravitinos, typical soft leptogenesis temperatures
are sensibly lower, and can accordingly relax this tension.  However,
soft leptogenesis is plagued by the problem of a congenital low
efficiency, that is related to the cancellation between the
asymmetries produced in fermions and bosons carrying lepton number. As
we have discussed in length, this cancellation becomes almost exact in
the zero temperature limit.  Eventually, finite temperature
corrections, that break supersymmetry and spoil the cancellation
between the scalar and fermion CP asymmetries, can rescue soft
leptogenesis from a complete failure.

It should be stressed at this point that the fact that lepton number
$L$ commutes with supersymmetric transformations, that is that scalar
and fermionic members of a supermultiplet have the same lepton number,
plays a crucial role in enforcing the aforementioned CP asymmetry
cancellation.

In this paper we have pointed out that in the temperature regime
quantified by \Eqn{eq:Tgmu}, in which all reactions that depend on the
soft gaugino masses do not occur, the early Universe effective theory
includes a new $R$-symmetry. In soft leptogenesis, this $R$-symmetry
is violated in the out of equilibrium interactions of neutrinos and
sneutrinos. In particular, $R$-number CP asymmetries in heavy
sneutrino decays can be defined, and constitute  important
quantities.  In fact, given that $R$-symmetries do not commute with
supersymmetry transformations, it is hardly surprising that 
no cancellation occurs between the $R$-number CP asymmetries for scalars
and fermions.  For this reason, a sizable density asymmetry for the
$R$ charge can develop in the thermal bath, and this asymmetry turns
out to be the main responsible for the generation of the baryon
asymmetry.

To keep higgsinos sufficiently light, in supersymmetry one needs to
assume $\muh \sim m_{\tilde g}$, and thus when the gaugino masses are
set to zero, one must set $\muh \to 0$ as well.  In this limit the
effective theory acquires another quasi-conserved global symmetry,
that is a $U(1)_{PQ}$ symmetry of the Peccei-Quinn type. $PQ$ is also
violated in sneutrino interactions and thus it also has an associated
CP asymmetry.  However, since $U(1)_{PQ}$ is a bosonic symmetry that
commutes with supersymmetry, the same cancellation between
fermion/boson $CP$ asymmetries occurring for lepton number also occurs
for $PQ$. Accordingly, $PQ$ does not play an equivalently important
role in the generation of the baryon asymmetry.

In order to make more understandable the previous two remarks, let us
start from the beginning, by listing the relevant global symmetries of
the effective theory. For simplicity we concentrate on Case~I
($h_{e,d}$ Yukawa equilibrium).  Neglecting lepton flavour, that is
irrelevant for the present discussion, these symmetries are:
$L,\,R,\,PQ,\,B$ and $\raise 2pt\hbox{$\chi$}_{u^c_L}$.  The first
three $L,\,R,\,PQ$ are violated perturbatively in the interactions of
the heavy sneutrinos, and all five symmetries are violated by
non-perturbative sphaleron processes. In this paper, in carrying out
our analysis, we have first identified the anomaly free combinations
of the five charges, that are $B-L$, $R_B$ and $R_\chi$, and then we
have written down the BE to describe their evolution. Here, we want to
sketch a different procedure. We first write  a set of evolution
equations for the five anomalous charges, that  have the form:
\begin{equation}
  \label{eq:calQ}
  \dot Y_{\Delta_{\cal Q}} = {\cal S}_{\Delta_{\cal Q}} +
 {\cal G}_{\Delta_{\cal Q}} + {\cal G}^{NP}_{\Delta_{\cal Q}}\,.
\end{equation}
In this equation ${\cal S}$ represent the source term for $
Y_{\Delta}$, ${\cal G}$ is the (s)neutrino-related washouts with all
density-asymmetries and signs absorbed, and ${\cal G}^{NP}$ represents
the non-perturbative EW and/or QCD sphaleron reactions that violate
$\Delta_{\cal Q}$.  The latter are reactions of type (i), that is fast
processes, that eventually will be convenient to eliminate in favor 
of chemical equilibrium conditions.  Now, given that $B$ and $\raise
2pt\hbox{$\chi$}_{u^c_L}$ are good symmetries at the perturbative
level, they have no CP-violating source term and ${\cal
  S}_{\Delta_{B}},\, {\cal S}_{\Delta_{\chi}}=0$ (they also do not
have perturbative washouts, and ${\cal G}_{\Delta_{B}}$ ${\cal
  G}_{\Delta_{\chi}}=0$ too). The only source terms thus are ${\cal
  S}_{\Delta_{L}},\, {\cal S}_{\Delta_{PQ}}$ and ${\cal
  S}_{\Delta_{R}}$. However, as we already know, in the $T\to 0$
limit, for ${\cal S}_{\Delta_{L}}$ we have a cancellation between the
fermion and scalar contributions: ${\cal S}^f_{\Delta_{L}}+ {\cal
  S}^s_{\Delta_{L}} \to 0$.  This straightforwardly implies that
${\cal S}^f_{\Delta_{PQ}}+ {\cal S}^s_{\Delta_{PQ}} \to 0$ too, since
the sneutrino processes contributing to the CP asymmetry for $PQ$ are
the same than for $L$: they are simply multiplied by the appropriate
$PQ$ charge that is, however, the same for fermion and scalar final
states. For the $R$ charge we have instead ${\cal S}_{\Delta_{R}}
\propto R_f\cdot {\cal S}^f_{\Delta_{L}}+ R_s\cdot {\cal
  S}^s_{\Delta_{L}}$, where $R_{f,s}$ are respectively the overall
$R$-charges of the fermion and boson {\it two particle} final state,
and thus satisfy $R_{s}=R_f+2$.  We then straightforwardly obtain that
in the $T\to 0$ limit the $R$-charge source term does not vanish, and
is given by ${\cal S}_{\Delta_{R}}\to 2\, {\cal S}^s_{\Delta_{L}}$.
Fast in-equilibrium sphaleron processes enforce equilibrium conditions
between particle densities carrying $R$ charge, and those carrying a
$B$ and $L$ numbers and, as a result, eventually baryon and lepton
asymmetries roughly of the same order than the $R$ charge-asymmetry
develop.  Eventually, with the decreasing of the temperature, gaugino
mass related reactions will start occurring with in-equilibrium rates
erasing any asymmetry in the $R$ charge.  It is important to notice
that when the $R$-symmetry gets explicitly broken, generalized EW
sphalerons reduce to the standard EW sphalerons and sphaleron induced
multi-fermion operators decouple from gauginos,\footnote{We are
  concentrating here on the role and fate of the $R$-symmetry.
  However, given that eventually also the $PQ$ symmetry gets
  explicitly broken, higgsinos decouple from sphalerons as well.}  
and reduce to their standard $B+L$ violating form. Since gaugino mass
reactions as well as all other MSSM processes conserve $B-L$, the
asymmetry initially generated through $R$-genesis will remain
unaffected.

Now that we have identified where the large density asymmetries come
from, we can complete our  procedure by constructing
suitable linear combinations of the five equations \eqref{eq:calQ} for
which the sphaleron terms ${\cal G}^{NP}$ cancel out. Since there are
only two such terms, ${\cal G}^{NP}_{EW}$ and ${\cal G}^{NP}_{QCD}$,
we can construct three linear combinations in which only processes of
type (iii) enter.  These are the BE equations for the three anomaly
free charges that have been discussed at length in
Section~\ref{sec:symmetries}.  The equilibrium conditions enforced by
${\cal G}^{NP}_{EW}$ and ${\cal G}^{NP}_{QCD}$ have to be imposed on
the system, and to obtain the BE in closed form, the various
density-asymmetries appearing in the washout terms ${\cal G}$ must be
rotated into the densities of the anomaly free charges by means of the
appropriate $A$ matrix.

In this paper, we have not formulated possible alternative effective
theories in which for example only $\muh = 0$ is set to zero, that
would correspond to an $U(1)_{PQ}$-theory, or the alternative case of
having just an $U(1)_R$-theory.  
However, we have written down a set of BE that are sufficiently
general to allow exploring numerically the outcome of such scenarios.
The corresponding results are resumed in Figure~\ref{fig:mg-eta}, and
confirm the crucial role played by the
$R$ symmetry.  In contrast, the effects ascribable to the new $PQ$
symmetry arising in the $\muh \to 0$ limit, that as we have seen are
not related with any new large CP violating source, remain of the
typical size of spectator effects.

In conclusion, supersymmetry offers different ways to explain the
cosmic matter-antimatter asymmetry.  The asymmetry could be directly
generated in baryon number since, although severely constrained, EW
baryogenesis has not been ruled out yet.  Alternatively, the asymmetry
could be initially generated in lepton number, through supersymmetric
leptogenesis~\cite{ournse} or through soft-leptogenesis if it occurs
below $T\sim 10^7\,$GeV.  The main finding of our paper is that there
is also a third, previously unnoticed, possibility.  That is that the
asymmetry can be first generated in the new $R$ charge that appears in
the effective theory for supersymmetry when the Universe temperature
is above $T\sim 10^7\,$GeV, and then  transferred to baryons via
generalized EW sphalerons.


\acknowledgments

This work is supported by  USA-NSF grant PHY-0653342 and by
Spanish  grants from MICINN 2007-66665-C02-01, the
INFN-MICINN agreement program ACI2009-1038,  consolider-ingenio 2010
program  CSD-2008-0037 and by CUR Generalitat de Catalunya grant
2009SGR502.


\bigskip


\appendix
\section{Thermal factors}
\label{Appendix-A}

In terms of the dimensionless evolution parameter $z=M/T$ 
the thermal factors appearing in the expressions of the decay CP 
asymmetries~\Eqns{eq:CP_asymresf}{eq:CP_asymint}  read:   
\begin{equation}
\Delta_{s,f}(z)  =  \frac{c^{s,f}(z)}{c^{s}(z) + c^{f}(z)}, 
\label{eq:thermal_factor}
\end{equation}
where, in the approximation in which $\widetilde N_\pm$ decay at rest,
\bea 
c^f(z) &=&(1-x_{\ell} -x_{\widetilde
H_u})\lambda(1,x_{\ell},x_{\widetilde H_u}) \left[ 1-\fLeq\right]
\left[ 1-\fht\right],
\label{cfeq}\\
c^s(z)&=&\lambda(1,x_{H_u},x_{\widetilde{\ell}}) \left[ 1+\fh\right]
\left[ 1+\fLteq\right],
\label{cbeq}
\eea 
with 
\bea
x_a(z) &=&\frac{m_a(z)^2}{M^2}\,,\\
\label{eq:xa}
\lambda(1,x,y)&=&\sqrt{(1+x-y)^2-4x}\,. 
\label{eq:lambda}
\eea 
The  Bose-Einstein $(s)$ and Fermi-Dirac $(f)$  equilibrium distributions
are:
\bea
f^{eq}_{s}&=&\frac{1}{e^{z\varepsilon_{\!s}}-1}\,,
\qquad\qquad
s=\tilde{\ell},H_u\,,
\label{eq:fHeq}\\
f^{eq}_{f}&=& \frac{1}{e^{z\varepsilon_{\!f}}+1},
\qquad\qquad  f=\ell,\widetilde H_u\,, 
\label{eq:fheq}
\eea 
where
\bea 
\varepsilon_{\ell,\widetilde H_u}&=&\frac{1}{2}
(1+x_{\ell,\widetilde H_u}- x_{\widetilde H_u,\ell}), \\
\varepsilon_{\widetilde{\ell},H_u}&=&
\frac{1}{2} (1+x_{\widetilde{\ell},H_u}-
x_{H_u,\widetilde{\ell}}).
\eea 
Finally, the thermal masses for the relevant scalar and fermion 
particle species are~\cite{thermal}: 
\bea 
x_{H_u}=2\, x_{\widetilde
  H_u}&=& \frac{1}{z^2}\left(\frac{3}{8}g_2^2+\frac{1}{8}g_Y^2
  +\frac{3}{4}\lambda_t^2\right)\; ,\\
x_{\widetilde{\ell}}\ =\ 2\, x_\ell&=&
\frac{1}{z^2}\left(\frac{3}{8}g_2^2+\frac{1}{8}g_Y^2 \right)\; , 
\eea 
where $g_2$ and $g_Y$ are the $SU(2)$ and $U(1)$ gauge couplings, and
$\lambda_t$ is the top Yukawa coupling, renormalized at the
appropriate energy scale.


\section{Boltzmann Equations}
\label{Appendix-B}

In this Appendix we present the Boltzmann equations that must be used
for numerical studies of soft leptogenesis when the heavy sneutrino
masses satisfy the condition~\Eqn{eq:NSE}. We also include the SE
reactions $\gamma^{\rm eff}_{\tilde g}$ and $\gamma^{\rm
  eff}_{\submuh}$ defined in \Eqn{eq:approxSE}, that extend the
validity of our BE to all temperatures.

The Boltzmann equations which describe the evolution of RH neutrino
and sneutrino densities are:
\begin{eqnarray}
\dot Y_{N} & \!= \! & -\left(\frac{Y_{N}}{Y_{N}^{eq}}
-1\right)\left(\gamma_{N}+4\gamma_{t}^{(0)}+4\gamma_{t}^{(1)}
+4\gamma_{t}^{(2)}+2\gamma_{t}^{(3)}+4\gamma_{t}^{(4)}\right),
\label{eq:BE_N}
\qquad\qquad
\mbox{\hspace{1.2cm} \phantom{i}} \\
\label{eq:BE_tildeN}
\dot Y_{\widetilde{N}} & \! =\! & 
-\left(\frac{Y_{\widetilde{N}}}{Y_{\widetilde{N}_+}^{eq}}-2\right) 
\left(\frac{\gamma_{\widetilde{N}}}{2}+3\gamma_{22}
+2\gamma_{t}^{\left(5\right)}+
2\gamma_{t}^{\left(6\right)}+2\gamma_{t}^{\left(7\right)}+
\gamma_{t}^{\left(8\right)}+2\gamma_{t}^{\left(9\right)}\right),
\end{eqnarray}
where the time derivative is defined as $\dot Y= sHz\frac{dY}{dz}$,
$s$ is the entropy density, and $H=H(z)$ is the Hubble parameter.  We
have defined $Y_{\widetilde N} \equiv Y_{\widetilde N_+} +
Y_{\widetilde N_-}\,$, and the reaction rates $\gamma$ without a
flavour index $\alpha$ are always understood to be summed over all
flavours. For the evolution of the flavour charges $Y_{\Delta_\alpha}$
we have
\begin{eqnarray}
\dot Y_{\Delta_\alpha} &=& -\left(E_\alpha+\widetilde E_\alpha\right),
\label{eq:BE_Delta_alpha} 
\end{eqnarray}
where
\begin{eqnarray}
E_\alpha & = & \epsilon_f^\alpha\left(z\right)
\frac{\gamma_{\widetilde{N}}}{2}
\left(\frac{Y_{\widetilde{N}}}{Y_{\widetilde{N}_+}^{eq}}-2\right)
-\frac{\gamma_{\widetilde{N}}^{f,\alpha}}{2}
\left({\cal Y}_{\Delta\ell_\alpha}+{\cal Y}_{\Delta \widetilde H_u}\right)
-\frac{1}{4}\gamma_{N}^\alpha\left({\cal Y}_{\Delta\ell_\alpha}+
{\cal Y}_{\Delta H_u}
\right)
\nonumber \\
 &  & -\left(\gamma_{t}^{\left(3\right)\alpha}\frac{Y_{N}}{Y_{N}^{eq}}+
2\gamma_{t}^{\left(4\right)\alpha}+2\gamma_{t}^{\left(6\right)\alpha}+
2\gamma_{t}^{\left(7\right)\alpha}+\gamma_{t}^{\left(5\right)\alpha}
\frac{Y_{\widetilde{N}}}{Y_{\widetilde{N}_+}^{eq}}\right)
{\cal Y}_{\Delta\ell_\alpha}
\nonumber \\
 &  & -\left(\gamma_{t}^{\left(3\right)\alpha}+\gamma_{t}^{\left(4\right)\alpha}
+\gamma_{t}^{\left(4\right)\alpha}\frac{Y_{N}}{Y_{N}^{eq}}+
\gamma_{t}^{\left(5\right)\alpha}+\gamma_{t}^{\left(6\right)\alpha}+
\frac{1}{2}\gamma_{t}^{\left(7\right)\alpha}
\frac{Y_{\widetilde{N}}}{Y_{\widetilde{N}_+}^{eq}}\right)
{\cal Y}_{\Delta H_u}
\nonumber \\
 &  & -\left(\gamma_{t}^{\left(5\right)k}+\gamma_{t}^{\left(7\right)k}+
\frac{1}{2}\gamma_{t}^{\left(6\right)k}
\frac{Y_{\widetilde{N}}}{Y_{\widetilde{N}_+}^{eq}}\right)
\left(2{\cal Y}_{\Delta \widetilde H_u}-{\cal Y}_{\Delta H_u}\right)
\nonumber \\
 &  & + \gamma_{\tilde g}^{\rm eff}
\left({\cal Y}_{\Delta \widetilde \ell_\alpha}-{\cal Y}_{\Delta\ell_\alpha}\right),
\label{eq:BE_Delta_lep}
\end{eqnarray}
and 
\begin{eqnarray}
\widetilde E_\alpha & = & 
\epsilon_s^\alpha\left(z\right)\frac{\gamma_{\widetilde{N}}}{2}
\left(\frac{Y_{\widetilde{N}}}{Y_{\widetilde{N}_+}^{eq}}-2\right)
-\frac{\gamma_{\widetilde{N}}^{s,\alpha}}{2}
\left({\cal Y}_{\Delta \widetilde \ell_\alpha}+{\cal Y}_{\Delta H_u}\right)
-\frac{1}{4}\gamma_{N}^\alpha\left({\cal Y}_{\Delta \widetilde \ell_\alpha}
+{\cal Y}_{\Delta \widetilde H_u}\right)
\nonumber \\
&  & -\left(
\frac{1}{2}\gamma_{22}^{\alpha}\frac{Y_{\widetilde{N}}}{Y_{\widetilde{N}_+}^{eq}}
+2\gamma_{22}^{\alpha}\right)\left({\cal Y}_{\Delta\widetilde{\ell}_\alpha}
+2{\cal Y}_{\Delta\widetilde{H}_u}-{\cal Y}_{\Delta H_u}
\right)\nonumber \\
&  & -\left(2\gamma_{t}^{\left(0\right)\alpha}\frac{Y_{N}}{Y_{N}^{eq}}+
2\gamma_{t}^{\left(1\right)\alpha}+2\gamma_{t}^{\left(2\right)\alpha}+
\frac{1}{2}\gamma_{t}^{\left(8\right)\alpha}
\frac{Y_{\widetilde{N}}}{Y_{\widetilde{N}_+}^{eq}}+
2\gamma_{t}^{\left(9\right)k}\right){\cal Y}_{\Delta \widetilde \ell_\alpha}
\nonumber \\
 &  & -\left(\gamma_{t}^{\left(0\right)\alpha}+\gamma_{t}^{\left(1\right)\alpha}
\frac{Y_{N}}{Y_{N}^{eq}}+\gamma_{t}^{\left(8\right)\alpha}
+\gamma_{t}^{\left(9\right)\alpha}
+\frac{1}{2}\gamma_{t}^{\left(9\right)\alpha}
\frac{Y_{\widetilde{N}}}{Y_{\widetilde{N}_+}^{eq}}\right){\cal Y}_{\Delta H_u}
\nonumber \\
 &  & -\left(\gamma_{t}^{\left(0\right)\alpha}+\gamma_{t}^{\left(1\right)\alpha}+
\gamma_{t}^{\left(2\right)\alpha}\frac{Y_{N}}{Y_{N}^{eq}}\right)
\left(2{\cal Y}_{\Delta \widetilde H_u}-{\cal Y}_{\Delta H_u}\right)
\nonumber \\
 &  & -\gamma_{\tilde g}^{\rm eff}
\left({\cal Y}_{\Delta \widetilde \ell_\alpha}-{\cal
    Y}_{\Delta\ell_\alpha}\right). 
\label{eq:BE_Delta_slep}
\end{eqnarray}
The ${\cal Y}_\Delta$ appearing in these equations are defined
in~\Eqn{eq:calY}, while the SE reaction rate $\gamma_{\tilde g}^{\rm
  eff}$ has been defined in~\Eqn{eq:approxSE}.  For the
decay reaction densities we have:
\begin{eqnarray}
\gamma_{\widetilde{N}}^{s,\alpha} &=&
\gamma_{\widetilde{N}}^{f,\alpha}\left(1+\frac{A^2}{M^2}-\frac{AB}{M^2}\right), 
\nonumber \\
\gamma_{\widetilde{N}}^\alpha&\equiv&\gamma_{\widetilde{N}}^{f,\alpha}
+\gamma_{\widetilde{N}}^{s,\alpha},
\label{eq:decay_readen}
\end{eqnarray}
where $A$ and $B$ are taken to be real.  For values $M \sim 10^8\,$GeV
the higher order terms in the soft parameters can be safely neglected.

The scattering processes considered are
\begin{displaymath}
\begin{array}{|l|l|l|}
\hline
\rm Reaction & \Delta R_B & \Delta R_3 \\ \hline
\gamma_{22}^{\alpha} 
 \equiv  \gamma\left(\widetilde{N}_{\pm}\widetilde{\ell}_{\alpha}\leftrightarrow\widetilde{Q}\widetilde{u}^*\right)
=\gamma\left(\widetilde{N}_{\pm}\widetilde{Q}^{*}\leftrightarrow\widetilde{\ell}_{\alpha}^{*}\widetilde{u}^*\right)
=\gamma\left(\widetilde{N}_{\pm}\widetilde{u}\leftrightarrow\widetilde{\ell}_{\alpha}^{*}\widetilde{Q}\right)
& 0 & 1 \\
\gamma_{t}^{(0)\alpha} 
 \equiv  \gamma\left(N\widetilde{\ell}_{\alpha}\leftrightarrow Q\widetilde{u}^*\right)
=\gamma\left(N\widetilde{\ell}_{\alpha}\leftrightarrow\widetilde{Q}\overline{u}\right)
& -1 & 0 \\
\gamma_{t}^{(1)\alpha} 
\equiv  \gamma\left(N\overline{Q}\leftrightarrow\widetilde{\ell}_{\alpha}^{*}\widetilde{u}^*\right)
=\gamma\left(Nu\leftrightarrow\widetilde{\ell}_{\alpha}^{*}\widetilde{Q}\right)
& -1 & 0 \\
\gamma_{t}^{(2)\alpha} 
\equiv  \gamma\left(N\widetilde{u}\leftrightarrow\widetilde{\ell}_{\alpha}^{*}Q\right)
=\gamma\left(N\widetilde{Q}^{*}\leftrightarrow\widetilde{\ell}_{\alpha}^{*}\overline{u}\right)
& -1 & 0 \\
\gamma_{t}^{(3)\alpha} 
 \equiv  \gamma\left(N\ell_{\alpha}\leftrightarrow Q\overline{u}\right)
& -1 & 0 \\
\gamma_{t}^{(4)\alpha} 
\equiv  \gamma\left(Nu\leftrightarrow\overline{\ell_{\alpha}}Q\right)
=\gamma\left(N\overline{Q}\leftrightarrow\overline{\ell_{\alpha}}\overline{u}\right)
& -1 & 0  \\
\gamma_{t}^{(5)\alpha} 
 \equiv  \gamma\left(\widetilde{N}_{\pm}\ell_{\alpha}\leftrightarrow Q\widetilde{u}^*\right)
=\gamma\left(\widetilde{N}_{\pm}\ell_{\alpha}\leftrightarrow\widetilde{Q}\overline{u}\right)
& 0 & 1  \\
\gamma_{t}^{(6)\alpha} 
\equiv  \gamma\left(\widetilde{N}_{\pm}\widetilde{u}\leftrightarrow\overline{\ell_{\alpha}}Q\right)
=\gamma\left(\widetilde{N}_{\pm}\widetilde{Q}^{*}\leftrightarrow\overline{\ell_{\alpha}}\overline{u}\right)
& 0 & 1 \\
\gamma_{t}^{(7)\alpha} 
 \equiv  \gamma\left(\widetilde{N}_{\pm}\overline{Q}\leftrightarrow\overline{\ell_{\alpha}}\widetilde{u}^*\right)
=\gamma\left(\widetilde{N}_{\pm}u\leftrightarrow\overline{\ell_{\alpha}}\widetilde{Q}\right)
& 0 & 1  \\
\gamma_{t}^{(8)\alpha} 
\equiv  \gamma\left(\widetilde{N}_{\pm}\widetilde{\ell}_{\alpha}^{*}\leftrightarrow\overline{Q}u\right)
& 2 & 1  \\
\gamma_{t}^{(9)\alpha} 
\equiv  \gamma\left(\widetilde{N}_{\pm}Q\leftrightarrow\widetilde{\ell}_{\alpha}u\right)
=\gamma\left(\widetilde{N}_{\pm}\overline{u}\leftrightarrow\widetilde{\ell}_{\alpha}\overline{Q}\right) 
& 2 & 1 \\
\hline
\end{array}
\end{displaymath}
where for convenience we have listed the corresponding changes
of the R-charges in each process. 
The reduced cross sections for the processes listed above can be found 
in ref.~\cite{Plumacher:1997ru}.

The BE above do not include the CP asymmetries of top and stop
scatterings.  Strictly speaking, when scatterings are included, for
consistency one should include also the corresponding CP asymmetries.
However, in soft leptogenesis this cannot be done in a straightforward
way because thermal factors for the scattering CP asymmetries
constitute a new set of non trivial quantities. Fortunately, in the
strong washout regime for leptogenesis, the effects of CP asymmetries
in scattering have been found to be subdominant with respect to CP
asymmetries in decays~\cite{CPscatt}, and since in this paper we focus
precisely on strong washouts, neglecting the scattering CP asymmetries
is justified.

The BE for the evolution of  $R_B$ and $R_\chi$,
defined in~\Eqns{eq:RB}{eq:chiralup}, are:
\begin{eqnarray}
\dot Y_{\Delta_{R_B}} &=& \sum_\alpha 
\left(2\widetilde F_\alpha + F_\alpha\right)
-\gamma^{\rm eff}_{\widetilde g}\,{\cal Y}_{\Delta \widetilde g} ,
\label{eq:BE_Delta_RB} 
\\
 \dot Y_{\Delta_{R_\chi}} 
& = &\frac{1}{3} \sum_\alpha  
\left(\widetilde G_\alpha - G_\alpha\right)
-\frac{\gamma^{\rm eff}_{\widetilde g}}{3}\, {\cal Y}_{\Delta \widetilde g}
+\frac{\gamma^{\rm eff}_{\submuh}}{3}
\left({\cal Y}_{\Delta\widetilde H_u}+{\cal Y}_{\Delta\widetilde H_d}\right),
\label{eq:BE_Delta_R3} 
\end{eqnarray}
where  again the SE rates $\gamma^{\rm eff}_{\widetilde g}$
and $\gamma^{\rm eff}_{\submuh}$ have been also included.
$F_{\alpha}$ and $\widetilde F_{\alpha}$ are given by:
\begin{eqnarray}
F_{\alpha} & = & -\frac{1}{4}
\gamma_{N}^\alpha\left({\cal Y}_{\Delta\ell_\alpha}+{\cal Y}_{\Delta H_u}\right)
 \nonumber \\ &  & 
-\left(\gamma_{t}^{\left(3\right)\alpha}\frac{Y_{N}}{Y_{N}^{eq}}+
2\gamma_{t}^{\left(4\right)\alpha}\right)
{\cal Y}_{\Delta\ell_\alpha}
\nonumber \\
 &  & -\left(\gamma_{t}^{\left(3\right)\alpha}+\gamma_{t}^{\left(4\right)\alpha}
+\gamma_{t}^{\left(4\right)\alpha}\frac{Y_{N}}{Y_{N}^{eq}}\right)
{\cal Y}_{\Delta H_u},
\label{eq:BE_F}
\end{eqnarray}
 and 
\begin{eqnarray}
\widetilde F_{\alpha} & = & 
\epsilon_s^\alpha\left(z\right)\frac{\gamma_{\widetilde{N}}}{2}
\left(\frac{Y_{\widetilde{N}}}{Y_{\widetilde{N}_+}^{eq}}-2\right)
-\frac{\gamma_{\widetilde{N}}^{s,\alpha}}{2}
\left({\cal Y}_{\Delta \widetilde \ell_\alpha}+{\cal Y}_{\Delta H_u}\right)
-\frac{1}{8}\gamma_{N}^\alpha\left({\cal Y}_{\Delta \widetilde \ell_\alpha}
+{\cal Y}_{\Delta \widetilde H_u}\right)
 \nonumber \\ 
&  & -\left(\gamma_{t}^{\left(0\right)\alpha}\frac{Y_{N}}{Y_{N}^{eq}}+
\gamma_{t}^{\left(1\right)\alpha}+\gamma_{t}^{\left(2\right)\alpha}+
\frac{1}{2}\gamma_{t}^{\left(8\right)\alpha}
\frac{Y_{\widetilde{N}}}{Y_{\widetilde{N}_+}^{eq}}+
2\gamma_{t}^{\left(9\right)\alpha}\right){\cal Y}_{\Delta \widetilde \ell_\alpha}
\nonumber \\
 &  & -\left(\frac{1}{2}\gamma_{t}^{\left(0\right)\alpha}
+\frac{1}{2}\gamma_{t}^{\left(1\right)\alpha}
\frac{Y_{N}}{Y_{N}^{eq}}+\gamma_{t}^{\left(8\right)\alpha}
+\gamma_{t}^{\left(9\right)\alpha}
+\frac{1}{2}\gamma_{t}^{\left(9\right)\alpha}
\frac{Y_{\widetilde{N}}}{Y_{\widetilde{N}_+}^{eq}}\right){\cal Y}_{\Delta H_u}
\nonumber \\
 &  & -\frac{1}{2}\left(\gamma_{t}^{\left(0\right)\alpha}+\gamma_{t}^{\left(1\right)\alpha}+
\gamma_{t}^{\left(2\right)\alpha}\frac{Y_{N}}{Y_{N}^{eq}}\right)
\left(2{\cal Y}_{\Delta \widetilde H_u}-{\cal Y}_{\Delta H_u}\right).
\label{eq:BE_tF}
\end{eqnarray}
For $G_{\alpha}$ and $\widetilde G_{\alpha}$ we have:
\begin{eqnarray}
G_{\alpha} & = & \epsilon_f^\alpha\left(z\right)
\frac{\gamma_{\widetilde{N}}}{2}
\left(\frac{Y_{\widetilde{N}}}{Y_{\widetilde{N}_+}^{eq}}-2\right)
-\frac{\gamma_{\widetilde{N}}^{f,\alpha}}{2}
\left({\cal Y}_{\Delta\ell_\alpha}+{\cal Y}_{\Delta \widetilde H_u}\right)
\nonumber \\
 &  & -\left(2\gamma_{t}^{\left(6\right)\alpha}+
2\gamma_{t}^{\left(7\right)\alpha}+\gamma_{t}^{\left(5\right)\alpha}
\frac{Y_{\widetilde{N}}}{Y_{\widetilde{N}_+}^{eq}}\right)
{\cal Y}_{\Delta\ell_\alpha}
\nonumber \\
 &  & -\left(\gamma_{t}^{\left(5\right)\alpha}+\gamma_{t}^{\left(6\right)\alpha}+
\frac{1}{2}\gamma_{t}^{\left(7\right)\alpha}
\frac{Y_{\widetilde{N}}}{Y_{\widetilde{N}_+}^{eq}}\right)
{\cal Y}_{\Delta H_u}
\nonumber \\
 &  & -\left(\gamma_{t}^{\left(5\right)\alpha}+\gamma_{t}^{\left(7\right)k}+
\frac{1}{2}\gamma_{t}^{\left(6\right)\alpha}
\frac{Y_{\widetilde{N}}}{Y_{\widetilde{N}_+}^{eq}}\right)
\left(2{\cal Y}_{\Delta \widetilde H_u}-{\cal Y}_{\Delta H_u}\right),
\label{eq:BE_G}
\end{eqnarray}
and 
\begin{eqnarray}
\widetilde G_{\alpha} & = & 
\epsilon_s^\alpha\left(z\right)\frac{\gamma_{\widetilde{N}}}{2}
\left(\frac{Y_{\widetilde{N}}}{Y_{\widetilde{N}_+}^{eq}}-2\right)
-\frac{\gamma_{\widetilde{N}}^{s,\alpha}}{2}
\left({\cal Y}_{\Delta \widetilde \ell_\alpha}+{\cal Y}_{\Delta H_u}\right)
\nonumber \\
&  & 
+\left(\frac{1}{2}\gamma_{22}^{\alpha}\frac{Y_{\widetilde{N}}}{Y_{\widetilde{N}_+}^{eq}}
+2\gamma_{22}^{\alpha}\right)\left({\cal Y}_{\Delta\widetilde{\ell}_\alpha}
+2{\cal Y}_{\Delta\widetilde{H}_u}-{\cal Y}_{\Delta H_u}
\right)\nonumber \\
&  & -\left(\frac{1}{2}\gamma_{t}^{\left(8\right)\alpha}
\frac{Y_{\widetilde{N}}}{Y_{\widetilde{N}_+}^{eq}}+
2\gamma_{t}^{\left(9\right)\alpha}\right){\cal Y}_{\Delta \widetilde \ell_\alpha}
\nonumber \\
 &  & -\left(\gamma_{t}^{\left(8\right)\alpha}
+\gamma_{t}^{\left(9\right)\alpha}
+\frac{1}{2}\gamma_{t}^{\left(9\right)\alpha}
\frac{Y_{\widetilde{N}}}{Y_{\widetilde{N}_+}^{eq}}\right){\cal Y}_{\Delta H_u}.
\label{eq:BE_tG}
\end{eqnarray}

As we have explained, with the inclusion of  $\gamma^{\rm
  eff}_{\widetilde g}$ and $\gamma^{\rm eff}_{\submuh}$ our
BE are valid at all temperatures. To verify this, we have compared the
results obtained with the complete BE given above, with what is
obtained by integrating the set of BE specific for the SE regime, that
reduce to the equations for the neutrino and sneutrino abundances
\Eqn{eq:BE_N} and \Eqn{eq:BE_tildeN} plus the three equations for the
flavour charges \Eqn{eq:BE_Delta_alpha}.  Of course, one also has to
use the $A^{\ell}$ matrices and $C^{\widetilde H_{u}}$ vectors
appropriate for the SE limits of the two cases that we have been
studying (recalling also  that $A^{\widetilde\ell}=2 A^{\ell}$ and
$C^{H_{u}} = 2C^{\widetilde H_{u}}$).
%
%
For Case I of Section~\ref{sec:edin} we have:
\begin{eqnarray}
A^\ell=\frac{1}{9\times 237}\left(
\begin{array}{ccc}
-221 & 16  & 16\\
 16 & -221 &  16 \\
 16 &  16  & -221
\end{array}\right), 
&\;\;\;\;\;& 
C^{\tilde H_u}=\frac{-4}{237}\left(1,\;1,\;1\right), 
\label{eq:ACTin}
\end{eqnarray}
that, incidentally, coincides with the matrix given in~\cite{ournse}
for the case of all Yukawa couplings in equilibrium.  The matrix for
Case~II of Section~\ref{sec:edout} is given in~\cite{ournse}, and is
rewritten below for convenience:
\begin{eqnarray}
A^\ell=\frac{1}{3\times 2148}
\left(\begin{array}{ccc}
-906 &\ \,\, 120  &\ \,\, 120\\
\ \, 75 &-688 &\ \, 28\\
\ \, 75 &\ \,28  &-688 
\end{array}\right), &\;& C^{\tilde H_u}=
\frac{-1}{2148}\left(37,\;52,\;52\right).
\label{eq:ACTout}
\end{eqnarray}
%



\bigskip 

\begin{thebibliography}{10}
%
%

\bibitem{fy}
M. Fukujita and T. Yanagida,
{\it Baryogenesis Without Grand Unification}
{\em Phys. Lett.} {\bf B174} (1986) 45
%
%

%
\bibitem{leptoreview} 
For a comprehensive review see:   S.~Davidson, E.~Nardi and Y.~Nir,
{\it Leptogenesis,}
{\em Phys. Rept.} {\bf 466}, (2008) 105;  
[\href{http://arxiv.org/abs/0802.2962}{{\tt arXiv:0802.2962}}].  


\bibitem{Luty:1992un}
  M.~A.~Luty,
  {\it Baryogenesis via leptogenesis},
  {\em Phys.\ Rev.}\  {\bf D45}, 455-465 (1992).

\bibitem{Covi:1996wh}
  L.~Covi, E.~Roulet, F.~Vissani,
  {\it CP violating decays in leptogenesis scenarios},
  {\em Phys.\ Lett.}\  {\bf B384}, 169-174 (1996)
[\href{http://arxiv.org/abs/hep-ph/9605319}{{\tt hep-ph/9605319}}].

\bibitem{Buchmuller:1996pa}
  W.~Buchmuller, M.~Plumacher,
 {\it Baryon asymmetry and neutrino mixing},
  {\em Phys.\ Lett.}\  {\bf B389}, 73-77 (1996)
[\href{http://arxiv.org/abs/hep-ph/9608308}{{\tt hep-ph/9608308}}].



\bibitem{flavour1}
 A.~Abada, S.~Davidson, A.~Ibarra, F.~X.~Josse-Michaux, M.~Losada and 
 A.~Riotto,  
{\it Flavour matters in leptogenesis},
{\em JHEP} {\bf 0609}, 010 (2006)
[\href{http://arxiv.org/abs/hep-ph/0605281}{{\tt hep-ph/0605281}}].
%
\bibitem{flavour2}
 E.~Nardi, Y.~Nir, E.~Roulet and J.~Racker,
{\it The importance of flavor in leptogenesis,}
{\em JHEP} {\bf 0601}, 164 (2006),
[\href{http://arxiv.org/abs/hep-ph/0601084}{{\tt hep-ph/0601084}}].  
%

\bibitem{barbieri}
R.~Barbieri, P.~Creminelli, A.~Strumia and N.~Tetradis,
{\it Baryogenesis through leptogenesis,}
{\em Nucl.\ Phys.\ B} {\bf 575} (2000) 61
[\href{http://arxiv.org/abs/hep-ph/9911315}{{\tt hep-ph/9911315}}].
%

\bibitem{morozumi}
T.~Endoh, T.~Morozumi and Z.~h.~Xiong,
  Prog.\ Theor.\ Phys.\  {\bf 111}, 123 (2004)
[\href{http://arxiv.org/abs/hep-ph/0308276}{{\tt hep-ph/0308276}}];
T.~Fujihara, S.~Kaneko, S.~Kang, D.~Kimura, T.~Morozumi and M.~Tanimoto,
  Phys.\ Rev.\ D {\bf 72}, 016006 (2005)
[\href{http://arxiv.org/abs/hep-ph/0505076}{{\tt hep-ph/0505076}}].

%
%

 \bibitem{oscar}
 O.~Vives,
{\it Flavoured leptogenesis: A successful thermal leptogenesis with N(1)  
  mass below 10**8-GeV},
  Phys.\ Rev.\ D {\bf 73} (2006) 073006
[\href{http://arxiv.org/abs/hep-ph/0512160}{{\tt hep-ph/0512160}}].
%

\bibitem{spectator2}
  E.~Nardi, Y.~Nir, J.~Racker and E.~Roulet,
{\it On Higgs and sphaleron effects during the leptogenesis era,}
{\em JHEP} {\bf 0601}, 068 (2006) 
[\href{http://arxiv.org/abs/hep-ph/0512052}{{\tt hep-ph/0512052}}].
%
\bibitem{spectator1}
W.~Buchmuller and M.~Plumacher,
{\it Spectator processes and baryogenesis},
{\em Phys.\ Lett. } {\bf B511}, 74 (2001)
[\href{http://arxiv.org/abs/hep-ph/0104189}{{\tt hep-ph/0104189}}].

\bibitem{ournse}
  C.~S.~Fong, M.~C.~Gonzalez-Garcia, E.~Nardi and J.~Racker,
  {\it Supersymmetric Leptogenesis},
{\em JCAP} {\bf 1012}, 013 (2010) 
[\href{http://arxiv.org/abs/1009.0003}{{\tt arXiv/1009.0003}}].  

%
\bibitem{Plumacher:1997ru}
  M.~Plumacher,
{\it Baryon asymmetry, neutrino mixing and supersymmetric SO(10)  
unification},
{\em Nucl.\ Phys.\  B} {\bf 530}, 207 (1998)
[\href{http://arxiv.org/abs/hep-ph/9704231}{{\tt hep-ph/9704231}}].  
%
%
\bibitem{thermal}
 G.~F.~Giudice, A.~Notari, M.~Raidal, A.~Riotto and A.~Strumia,
{\it Towards a complete theory of thermal leptogenesis in the SM and MSSM},
 {\em  Nucl.\ Phys.\ B} {\bf 685} (2004) 89,
[\href{http://arxiv.org/abs/hep-ph/0310123}{{\tt hep-ph/0310123}}].
%
%
\bibitem{Ibanez:1992aj}
L.~E.~Ibanez and F.~Quevedo,
{\it Supersymmetry Protects The Primordial Baryon Asymmetry},
{\em Phys.\ Lett. }  {\bf B283}, 261 (1992)
[\href{http://arxiv.org/abs/hep-ph/9204205}{{\tt hep-ph/9204205}}].  
%
\bibitem{Boubekeur:2002jn}
  L.~Boubekeur,
  {\it Leptogenesis at low scale},
[\href{http://arxiv.org/abs/hep-ph/0208003}{{\tt hep-ph/0208003}}].  
%
\bibitem{soft1}
Y.~Grossman, T.~Kashti, Y.~Nir and E.~Roulet,
{\it Leptogenesis from supersymmetry breaking,}
{\em Phys.\ Rev.\ Lett.}\  {\bf 91} (2003) 251801
[\href{http://arxiv.org/abs/hep-ph/0307081}{{\tt hep-ph/0307081}}].  
%
\bibitem{soft2}
G.~D'Ambrosio, G.~F.~Giudice and M.~Raidal,
{\it Soft leptogenesis},
{\em Phys.\ Lett.}\  {\bf B575}, 75 (2003)
[\href{http://arxiv.org/abs/hep-ph/0308031}{{\tt hep-ph/0308031}}].  
%
\bibitem{PFL1}
  D.~Aristizabal Sierra, M.~Losada and E.~Nardi,
 {\it Variations on leptogenesis}, 
  Phys.\ Lett.\  B {\bf 659}, 328 (2008)
[\href{http://arxiv.org/abs/0705.1489}{{\tt arXiv/0705.1489}}].  
%
\bibitem{PFL2}
  D.~A.~Sierra, L.~A.~Munoz and E.~Nardi,
  {\it Purely Flavored Leptogenesis},
Phys.\ Rev.\  D {\bf 80}, 016007 (2009)
[\href{http://arxiv.org/abs/0904.3043}{{\tt arXiv/0904.3043}}];  
%
  {\it Implications of an additional scale on leptogenesis}, 
 J.\ Phys.\ Conf.\ Ser.\  {\bf 171}, 012078 (2009)
[\href{http://arxiv.org/abs/0904.3052}{{\tt arXiv/0904.3052}}].  
%
\bibitem{lfe}
  D.~Aristizabal Sierra, M.~Losada and E.~Nardi,
{\it Lepton Flavor Equilibration and Leptogenesis},
{\em JCAP} {\bf 0912}, 015 (2009)
[\href{http://arxiv.org/abs/0905.0662}{{\tt arXiv/0905.0662}}].  
%
\bibitem{Antusch:2009gn}
  S.~Antusch, S.~Blanchet, M.~Blennow and E.~Fernandez-Martinez,
  JHEP {\bf 1001}, 017 (2010)
[\href{http://arxiv.org/abs/0910.5957}{{\tt arXiv/0910.5957}}].  
%
\bibitem{Fischler:1990gn}
  W.~Fischler, G.~F.~Giudice, R.~G.~Leigh and S.~Paban,
  {\it Constraints on the baryogenesis scale from neutrino masses}, 
  Phys.\ Lett.\  {\bf B258}, 45-48 (1991).
%
\bibitem{di}
  S.~Davidson and A.~Ibarra,
{\it A lower bound on the right-handed neutrino mass from leptogenesis},
 {\rm Phys.\ Lett. } {\bf B535} (2002) 25,
[\href{http://arxiv.org/abs/hep-ph/0202239}{{\tt hep-ph/0202239}}].  
%
\bibitem{Mbound}
W.~Buchmuller, P.~Di Bari and M.~Plumacher,
{\it Cosmic microwave background, matter-antimatter asymmetry and 
neutrino  masses},
Nucl.\ Phys.\ B {\bf 643} (2002) 367 
[\href{http://arxiv.org/abs/hep-ph/0205349}{{\tt hep-ph/0205349}}].
J.~R.~Ellis and M.~Raidal,
{\it Leptogenesis and the violation of 
lepton number and CP at low energies},
Nucl.\ Phys.\ B {\bf 643} (2002) 229 
[\href{http://arxiv.org/abs/hep-ph/0206174}{{\tt hep-ph/0206174}}].
%
\bibitem{flavour3}
  A.~Abada, S.~Davidson, F.~X.~Josse-Michaux, M.~Losada and A.~Riotto,
 {\it Flavour issues in leptogenesis},
{\em JCAP} {\bf 0604} (2006) 004
[\href{http://arxiv.org/abs/hep-ph/0601083}{{\tt hep-ph/0601083}}].  
%
 \bibitem{db2}
  S.~Blanchet and P.~Di Bari,
  {\it Flavor effects on leptogenesis predictions}, 
  JCAP {\bf 0703}, 018 (2007)
[\href{http://arxiv.org/abs/hep-ph/0607330}{{\tt hep-ph/0607330}}]; 
%

%
\bibitem{PU}
  A.~Pilaftsis and T.~E.~J.~Underwood,
  {\it Resonant leptogenesis},  
  Nucl.\ Phys.\  B {\bf 692} (2004) 303
[\href{http://arxiv.org/abs/hep-ph/0309342}{{\tt hep-ph/0309342}}];  
A.~Pilaftsis and T.~E.~J.~Underwood,
{\it Electroweak-scale resonant leptogenesis}, 
 Phys.\ Rev.\ D {\bf 72} (2005) 113001
[\href{http://arxiv.org/abs/hep-ph/0506107}{{\tt hep-ph/0506107}}];  
%
A.~Pilaftsis,
{\it Resonant tau leptogenesis with observable lepton number
  violation}, 
  Phys.\ Rev.\ Lett.\  {\bf 95}, 081602 (2005)
[\href{http://arxiv.org/abs/hep-ph/0408103}{{\tt hep-ph/0408103}}].  
%

\bibitem{lowtemp}
 E.~Ma, N.~Sahu and U.~Sarkar,
{\it Leptogenesis below the Davidson-Ibarra bound}, 
  J.\ Phys.\ G {\bf 32}, L65 (2006)
[\href{http://arxiv.org/abs/hep-ph/0603043}{{\tt hep-ph/0603043}}];  
%
  Y.~Farzan and J.~W.~F.~Valle,
{\it R-parity violation assisted thermal leptogenesis in the seesaw
  mechanism}, 
  Phys.\ Rev.\ Lett.\  {\bf 96}, 011601 (2006)
[\href{http://arxiv.org/abs/hep-ph/0509280}{{\tt hep-ph/0509280}}];  
%
  N.~Okada and O.~Seto,
{\it Thermal leptogenesis in brane world cosmology},
  Phys.\ Rev.\  D {\bf 73}, 063505 (2006)
[\href{http://arxiv.org/abs/hep-ph/0507279}{{\tt hep-ph/0507279}}].  
%
\bibitem{Casas:2004gh}
  J.~A.~Casas, J.~R.~Espinosa, I.~Hidalgo,
 {\it 
Implications for new physics from fine-tuning arguments. 
1. Application to SUSY and seesaw cases}, 
  JHEP {\bf 0411}, 057 (2004) 
[\href{http://arxiv.org/abs/hep-ph/0410298}{{\tt hep-ph/0410298}}].
%
%
\bibitem{gravi} 
M.~Y.~Khlopov and A.~D.~Linde,
{\it Is It Easy To Save The Gravitino?},
{\em Phys.\ Lett. } {\bf B 138} (1984) 265;
J.~R.~Ellis, J.~E.~Kim and D.~V.~Nanopoulos,
{\it Cosmological Gravitino Regeneration And Decay}
{\em Phys.\ Lett. } {\bf B145} (1984) 181;
J.~R.~Ellis, D.~V.~Nanopoulos and S.~Sarkar,
{\it The Cosmology Of Decaying Gravitinos}
{\em Nucl.\ Phys.\ B} {\bf 259} (1985) 175;
T.~Moroi, H.~Murayama and M.~Yamaguchi,
{\it Cosmological constraints on the light stable gravitino},
{\em Phys.\ Lett. } B {\bf B303} (1993) 289;
M.~Kawasaki, K.~Kohri and T.~Moroi,
{\it Hadronic decay of late-decaying particles and big-bang nucleosynthesis,}
{\em Phys.\ Lett.} {\bf B 625} (2005) 7
[\href{http://arxiv.org/abs/astro-ph/0402490}{{\tt astro-ph/0402490}}].  
%
\bibitem{ourflasoft}
C.~S.~Fong and M.~C.~Gonzalez-Garcia,
{\it Flavoured Soft Leptogenesis},
{\em JHEP} {\bf 0806}, 076 (2008)
[\href{http://arxiv.org/abs/0804.4471}{{\tt arXiv/0804.4471}}].  
%
\bibitem{ourqbe}
 C.~S.~Fong and M.~C.~Gonzalez-Garcia,
  {\it On Quantum Effects in Soft Leptogenesis}, 
  JCAP {\bf 0808} (2008) 008
[\href{http://arxiv.org/abs/0806.3077}{{\tt arXiv/0806.3077}}].  
%
\bibitem{ourinvsoft}
 J.~Garayoa, M.~C.~Gonzalez-Garcia and N.~Rius,
  {\it Soft leptogenesis in the inverse seesaw model}, 
  JHEP {\bf 0702} (2007) 021
[\href{http://arxiv.org/abs/hep-ph/0611311}{{\tt hep-ph/0611311}}].  
%
\bibitem{softothers}
G.~D'Ambrosio, T.~Hambye, A.~Hektor, M.~Raidal and A.~Rossi,
  {\it Leptogenesis in the minimal supersymmetric triplet seesaw
    model}, 
  Phys.\ Lett.\ B {\bf 604} (2004) 199
[\href{http://arxiv.org/abs/hep-ph/0407312}{{\tt hep-ph/0407312}}];  
%
  M.~C.~Chen and K.~T.~Mahanthappa,
  {\it Lepton flavor violating decays, soft leptogenesis and SUSY
    SO(10)}, 
  Phys.\ Rev.\  D {\bf 70}, 113013 (2004)
[\href{http://arxiv.org/abs/hep-ph/0409096}{{\tt hep-ph/0409096}}];  
%
 Y.~Grossman, R.~Kitano and H.~Murayama,
  {\it Natural soft leptogenesis}, 
  JHEP {\bf 0506}, 058 (2005)
[\href{http://arxiv.org/abs/hep-ph/0504160}{{\tt hep-ph/0504160}}];  
%
 E.~J.~Chun and S.~Scopel,
  {\it Soft leptogenesis in Higgs triplet model}, 
  Phys.\ Lett.\  B {\bf 636}, 278 (2006)
[\href{http://arxiv.org/abs/hep-ph/0510170}{{\tt hep-ph/0510170}}];  
%
A.~D.~Medina and C.~E.~M.~Wagner,
  {\it Soft leptogenesis in warped extra dimensions}, 
  JHEP {\bf 0612}, 037 (2006)
[\href{http://arxiv.org/abs/hep-ph/0609052}{{\tt hep-ph/0609052}}];  
%
  E.~J.~Chun and L.~Velasco-Sevilla,
   {\it SO(10) unified models and soft leptogenesis}, 
  JHEP {\bf 0708}, 075 (2007)
[\href{http://arxiv.org/abs/hep-ph/0702039}{{\tt hep-ph/0702039}}].  
%
\bibitem{soft3}
  Y.~Grossman, T.~Kashti, Y.~Nir and E.~Roulet,
  {\it New ways to soft leptogenesis},
  JHEP {\bf 0411} (2004) 080
[\href{http://arxiv.org/abs/hep-ph/0407063}{{\tt hep-ph/0407063}}].  
%
\bibitem{ourgaugino}
 C.~S.~Fong and M.~C.~Gonzalez-Garcia,
  {\it On Gaugino Contributions to Soft Leptogenesis}, 
  JHEP {\bf 0903} (2009) 073
[\href{http://arxiv.org/abs/0901.0008}{{\tt arXiv/0901.0008}}].
%
\bibitem{flavourothers}
%
S.~Pascoli, S.~T.~Petcov and A.~Riotto,
{\it Connecting low energy leptonic CP-violation to leptogenesis}, 
 Phys.\ Rev.\  D {\bf 75}, 083511 (2007)
[\href{http://arxiv.org/abs/hep-ph/0609125}{{\tt hep-ph/0609125}}];
%
  G.~C.~Branco, R.~Gonzalez Felipe and F.~R.~Joaquim,
{\it A new bridge between leptonic CP violation and leptogenesis}, 
  Phys.\ Lett.\  B {\bf 645} (2007) 432
[\href{http://arxiv.org/abs/hep-ph/0609297}{{\tt hep-ph/0609297}}];
%
S.~Antusch and A.~M.~Teixeira,
{\it Towards constraints on the SUSY seesaw from flavour-dependent
leptogenesis}, 
 JCAP {\bf 0702}, 024 (2007)
[\href{http://arxiv.org/abs/hep-ph/0611232}{{\tt hep-ph/0611232}}]; 
%
S.~Pascoli, S.~T.~Petcov and A.~Riotto,
{\it Leptogenesis and low energy CP violation in neutrino physics}, 
 Nucl.\ Phys.\  B {\bf 774}, 1 (2007)
[\href{http://arxiv.org/abs/hep-ph/0611338}{{\tt hep-ph/0611338}}];
  G.~Engelhard, Y.~Grossman, E.~Nardi and Y.~Nir,
{\it The importance of $N_2$ leptogenesis},
{\em Phys.\ Rev.\ Lett.}\  {\bf 99}, 081802 (2007)
[\href{http://arxiv.org/abs/hep-ph/0612197}{{\tt hep-ph/0612187}}].  
%
%
%
\bibitem{riottoqbefla}
  V.~Cirigliano, A.~De Simone, G.~Isidori, I.~Masina and A.~Riotto,
  {\it Quantum Resonant Leptogenesis and Minimal Lepton Flavour
    Violation},    JCAP {\bf 0801} (2008) 004
[\href{http://arxiv.org/abs/0711.0778}{{\tt arXiv/0711.0778}}].
%
\bibitem{riottosc}
 A.~De Simone and A.~Riotto,
{\it On the impact of flavour oscillations in leptogenesis}, 
 JCAP {\bf 0702} (2007) 005
[\href{http://arxiv.org/abs/hep-ph/0611357}{{\tt hep-ph/0611357}}]; 
%
S.~Blanchet, P.~Di Bari and G.~G.~Raffelt,
{\it Quantum Zeno effect and the impact of flavour in leptogenesis}, 
 JCAP {\bf 0703}, 012 (2007)
[\href{http://arxiv.org/abs/hep-ph/0611337}{{\tt hep-ph/0611337}}].
%
\bibitem{oursoftnu}
C.~S.~Fong, M.~C.~Gonzalez-Garcia, E.~Nardi and J.~Racker,
{\it Flavoured soft leptogenesis and natural values of the B term},
{\em JHEP} {\bf 1007}, 001 (2010)
[\href{http://arxiv.org/abs/1004.5125}{{\tt arXiv/1004.5125}}].  
%
%
\bibitem{ha90} 
J.~A.~Harvey and M.~S.~Turner,
{\it Cosmological Baryon And Lepton Number In The Presence Of Electroweak
Fermion Number Violation},
{\em Phys.\ Rev.\ D} {\bf 42}, 3344 (1990).
%
\bibitem{eR-equilibrium}
J.~M.~Cline, K.~Kainulainen and K.~A.~Olive,
{\it On the erasure and regeneration of the primordial baryon asymmetry by
 sphalerons},
{\em Phys.\ Rev.\ Lett.}\  {\bf 71}, 2372 (1993)
[\href{http://arxiv.org/abs/hep-ph/9304321}{{\tt hep-ph/9304321}}];  
J.~M.~Cline, K.~Kainulainen and K.~A.~Olive,
{\it Protecting the primordial baryon asymmetry from erasure by sphalerons},
{\em Phys.\ Rev.\  D} {\bf 49}, 6394 (1994)
[\href{http://arxiv.org/abs/hep-ph/9401208}{{\tt hep-ph/9401208}}].  
%
\bibitem{Keung:1983nq}
  W.~Y.~Keung and L.~Littenberg,
  {\it Test Of Supersymmetry In E- E- Collision},
  Phys.\ Rev.\  D {\bf 28}, 1067 (1983).
%
\bibitem{CPscatt}
  E.~Nardi, J.~Racker and E.~Roulet,
{\it CP violation in scatterings, three body processes and the Boltzmann
equations for leptogenesis},
{\em JHEP} {\bf 0709}, 090 (2007)
[\href{http://arxiv.org/abs/0707.0378}{{\tt arXiv/0707.0378}}];  
  C.~S.~Fong, M.~C.~Gonzalez-Garcia and J.~Racker,
  {\it CP Violation from Scatterings with Gauge Bosons in
    Leptogenesis}, 
[\href{http://arxiv.org/abs/1010.2209}{{\tt arXiv/1010.2209}}].  


\end{thebibliography}
\end{document}